\documentclass[preprintnumbers,aps,showpacs,nofootinbib,superscriptaddress,floatfix]{revtex4-1} 

\usepackage{graphicx}
\usepackage{amsmath}
\usepackage{amssymb}
\usepackage{graphicx,psfrag}
\usepackage{epsfig}
\usepackage{epstopdf}
\usepackage{afterpage}
\usepackage{mathrsfs}

\usepackage{bm}% bold math
\usepackage{setspace}
 
\usepackage{subfigure}
\usepackage{graphics,epsfig,color,latexsym}
\usepackage{graphicx}% Include figure files
\usepackage{dcolumn}% Align table columns on decimal point
\usepackage{bm}% bold math
\usepackage{epsfig}% Include figure files
\usepackage{wrapfig}
\usepackage[active]{srcltx} 
\usepackage{xspace}

\newcommand{\bgc}{\begin{center}}

\newcommand{\be}{\begin{equation}}
\newcommand{\ee}{\end{equation}}
\newcommand{\ben}{\begin{eqnarray}}
\newcommand{\een}{\end{eqnarray}}
\def\bea#1\eea{\begin{align}#1\end{align}}

\newcommand{\bef}{\begin{figure}[htb]\centering}
\newcommand{\eef}{\end{figure}}

\newcommand{\bfl}{\begin{flushleft}}
\newcommand{\efl}{\end{flushleft}}

\newcommand{\Php}{P_{h\perp}}

\newcommand{\enc}{\end{center}}

\newcommand{\bgi}{\begin{itemize}}
\newcommand{\eni}{\end{itemize}}

\begin{document}

%
% Title, Authors, Affiliations 
% ===========================
%
\preprint{JLAB-THY-14-1915}
\preprint{RBRC-1084}

\title{Left-right spin asymmetry in {\boldmath $\ell \, N^\uparrow \!\!\! \to h \, X$}}

\author{Leonard Gamberg}
\email{lpg10@psu.edu}
\affiliation{Division of Science, Penn State Berks, Reading, PA 19610, USA}
\author{Zhong-Bo Kang}
\email{zkang@lanl.gov}
\affiliation{Los Alamos National Laboratory, Theoretical Division, Los Alamos, NM 87545, USA}
\author{Andreas Metz}
\email{metza@temple.edu}
\affiliation{Department of Physics, Barton Hall, Temple University, Philadelphia, PA 19122, USA}
\author{Daniel Pitonyak}
\email{dpitonyak@quark.phy.bnl.gov}
\affiliation{RIKEN BNL Research Center, Brookhaven National Laboratory, Upton, NY 11973, USA}
\author{Alexei Prokudin}
\email{prokudin@jlab.org}
\affiliation{Jefferson Lab, 12000 Jefferson Avenue, Newport News, VA 23606, USA}

% Hadron-induced inclusive production with identified leptons, 
                % photons, or other nonhadronic particles (energy > 10 GeV)

%
% Abstract
% ===========================
%
\begin{abstract}
\noindent
We consider the inclusive production of hadrons in lepton-nucleon scattering.
For a transversely polarized nucleon this reaction shows a left-right azimuthal asymmetry, which we compute in twist-3 collinear factorization at leading order in perturbation theory. 
All non-perturbative parton correlators of the calculation are fixed through information from other hard processes.
Our results for the left-right asymmetry agree in sign with recent data for charged pion production from the HERMES Collaboration and from Jefferson Lab.
However, the magnitude of the computed asymmetries tends to be larger than the data.
Potential reasons for this outcome are identified.
We also give predictions for future experiments and highlight in particular the unique opportunities at an Electron Ion Collider.
\end{abstract}

\maketitle

%
% 1. Section: Introduction
% ===========================
%
\section{Introduction}
\noindent
In the present work we study inclusive production of hadrons in lepton-nucleon scattering, $\ell \, N \to h \, X$.
If the transverse momentum $P_{h\perp}$ of the final state hadron is sufficiently large, this process may be treated in perturbative Quantum Chromodynamics (QCD) and, therefore, can provide additional information about the parton structure of the nucleon.
Our focus here is on the left-right azimuthal asymmetry that can be defined if the nucleon is transversely polarized.
This asymmetry is similar to the transverse single-spin asymmetry $A_N$ which has already been studied extensively in hadronic collisions like $p^{\uparrow} p \to h \, X$ --- see Refs.~\cite{Adams:1991cs,Adler:2005in,Adare:2013ekj,Adare:2014qzo,Adams:2003fx,Abelev:2008qb,Adamczyk:2012xd,Arsene:2008mi,Bland:2013pkt,Efremov:1981sh,Efremov:1984ip,Qiu:1991pp,Qiu:1991wg,Qiu:1998ia,Kouvaris:2006zy,Kang:2011hk,Anselmino:1994tv,Anselmino:1998yz,Anselmino:2005sh,Anselmino:2012rq,Anselmino:2013rya,Kanazawa:2000hz,Eguchi:2006mc,Koike:2009ge,Kanazawa:2010au,Kanazawa:2011bg,Beppu:2013uda,Kang:2010zzb, Metz:2012ui, Metz:2012ct, Kanazawa:2014dca} for related experimental and theoretical work.
Recently, the HERMES Collaboration~\cite{Airapetian:2013bim} and the Jefferson Lab Hall A Collaboration~\cite{Allada:2013nsw} reported the first ever measurements of $A_N$ in lepton-nucleon scattering.
In general, one may expect that $A_N$ in this reaction could give new insight into the underlying mechanism of $A_N$ in hadronic collisions which is the subject of longstanding discussions.

We compute $A_N$ in collinear twist-3 factorization where it has two main components:
First, a twist-3 effect originates from the transversely polarized nucleon.
In that case the key non-perturbative entity is the so-called Qiu-Sterman (QS) function~\cite{Qiu:1991pp,Qiu:1991wg} --- a specific quark-gluon-quark correlator that has an intimate connection with the transverse momentum dependent (TMD) Sivers function~\cite{Sivers:1989cc,Sivers:1990fh}.
In a closely related previous work we have shown how the QS function can be studied through measuring $A_N$ for the process
$\ell \, p^{\uparrow} \to \textrm{jet} \, X$~\cite{Kang:2011jw}.
Second, a twist-3 effect also arises from parton fragmentation.
This contribution can be expressed by means of two independent fragmentation correlators~\cite{Yuan:2009dw,Metz:2012ct,Kanazawa:2013uia}, one of which is related to the Collins fragmentation function (FF)~\cite{Collins:1992kk}.
A first attempt to get a complete result for $A_N$ in $\ell \, p^{\uparrow} \to h \, X$ in the collinear twist-3 approach can be found in a conference proceeding~\cite{Koike:2002ti}.
Note that the same observable has also been computed in the so-called Generalized Parton Model (GPM), which uses TMD parton correlation functions~\cite{Anselmino:1999gd,Anselmino:2009pn,Anselmino:2014eza}. 

We fix all the non-perturbative parts of the analytical result for $A_N$ through available information from other hard scattering processes.
In particular, we take into account important input for the fragmentation correlators from a recent analysis of $A_N$ in $p^{\uparrow} p \to \pi \, X$~\cite{Kanazawa:2014dca}.
Our calculation agrees in sign with the data from HERMES~\cite{Airapetian:2013bim} and from Jefferson Lab~\cite{Allada:2013nsw}.
On the other hand, the results tend to be larger than the data.
As we discuss below in more detail, the most important reasons for this outcome   could be the underestimated error of our calculation,  and the impact from higher order corrections.
Such corrections can be expected to be very large for $\ell \, N \to h \, X$ in the kinematical region of the presently available data.
We therefore emphasize the need for a next-to-leading order (NLO) calculation of $A_N$ in order to explore to what extent this observable is theoretically under control. 
We also stress the importance of new experiments, in particular at a future Electron Ion Collider (EIC)~\cite{Boer:2011fh,Accardi:2011mz,Accardi:2012qut}.
With such a facility one could extend the measurements to (much) higher values of $P_{h\perp}$ where the perturbative expansion converges better.
Moreover, an EIC  would allow one for the first time
to explore the forward region of the nucleon in a lepton-nucleon reaction.
Note that it is precisely this forward region of the polarized nucleon where strikingly large asymmetries $A_N$ have been observed in $p^{\uparrow} p \to h \, X$~\cite{Adams:1991cs,Adare:2013ekj,Adams:2003fx,Abelev:2008qb,Adamczyk:2012xd,Arsene:2008mi}. 
 
Our paper is organized as follows. 
In Section \ref{sectionII} we present some details of the kinematics for $\ell \, N \to h \, X$ as well as our analytical results.
The numerical results are given in Section \ref{sectionIII}.
They include the comparison to existing data and predictions for future experiments.
 In Section \ref{sectionIII}  we also briefly compare our approach with the GPM.
The paper is summarized in Section \ref{sectionIV}.

%
% 2. Section: Kinematics and analytical results
% ===========================
%
\section{Kinematics and analytical results \label{sectionII}}
\noindent
Here we discuss some details of the kinematics and present the tree level results for the unpolarized and the spin-dependent cross section entering the definition of $A_N$. 
For the process under consideration 
\bea
\ell (l)  + N(P, S_P) \to h(P_h) + X \,,
\eea
$l$, $P$, and $P_h$ denote the momentum of the lepton, nucleon, and produced hadron, respectively, and $S_P$ is the spin vector of the nucleon. 
We use the momenta of the particles to fix a coordinate system according to $\hat{e}_z = \hat{P} = - \hat{l}$, $\hat{e}_x = \hat{P}_{h\perp}$, and $\hat{e}_y = \hat{e}_z \times \hat{e}_x$.
The Mandelstam variables for the scattering process are defined by
\begin{equation} \label{e:mandel_1}
S = (l + P)^2 \,, \qquad
T = (P - P_h)^2 \,, \qquad 
U = (l - P_h)^2 \,,
\end{equation}
while at the corresponding  partonic level one has
\begin{equation} \label{e:mandel_2}
\hat{s} = (l + k)^2 = x S\,, \qquad
\hat{t} = (k - p)^2 = \frac{x T}{z} \,, \qquad 
\hat{u} = (l - p)^2 = \frac{U}{z} \,,
\end{equation}
with $k$ characterizing the momentum of the active quark in the nucleon, and $p$ the momentum of the fragmenting quark. 
Neglecting parton transverse momenta one has $k = x P$ and $p = P_h / z$.

For the unpolarized lepton-nucleon collisions, the differential cross section at leading order (LO) is given by~\cite{Kang:2011jw}
\begin{equation}
P_h^0 \, \frac{d\sigma_{UU}} {d^3\vec{P}_h} = \frac{2\alpha_{\rm em}^2} {S} \,
\sum_q e_q^2 \int_{z_{\rm min}}^1 \frac{dz} {z^2} \, \frac{1} {S+T/z} \, \frac{1} {x} \, f_1^q(x) \, D_1^{h/q}(z) 
\bigg[ \frac{\hat{s}^2 + \hat{u}^2}{\hat{t}^2} \bigg] \,,
\label{e:lNhXUU}
\end{equation}
where $f_1^q$ is the unpolarized quark distribution, and $D_1^{h/q}$ is the unpolarized fragmentation function.
Here $z_{\rm min}= -(T+U)/S$, and $x$ can be determined from the on-shell condition $\hat s+ \hat t + \hat u=0$ in our LO formula as
\bea
x=-(U/z)/(S+T/z) \,.
\label{e:xdef}
\eea
We now turn to the spin-dependent cross section for the process $\ell \, N^{\uparrow}  \to h \, X$, that is,  an unpolarized lepton scattering off 
a transversely polarized nucleon. 
We work in the collinear factorization framework, in which this cross section is a twist-3 observable. 
The twist-3 effect can either come from the side of the parton distribution in the transversely polarized nucleon~\cite{Qiu:1991pp}, or from the side of the parton fragmentation into the final-state hadron~\cite{Yuan:2009dw,Kang:2010zzb,Metz:2012ct}. 
Calculations for such a twist-3 observable in collinear factorization have become standard, and details can be found in the literature --- see, e.g., Refs.~\cite{Qiu:1991pp,Qiu:1991wg,Qiu:1998ia,Kouvaris:2006zy,Kanazawa:2000hz,Eguchi:2006mc,Koike:2009ge,Kang:2010zzb,Metz:2012ct,Yuan:2009dw,Kanazawa:2013uia,Kang:2008qh,Kang:2008ih,Zhou:2009jm,Beppu:2010qn,Koike:2011mb,Liang:2012rb,Metz:2012fq,Hatta:2013wsa}. In particular, we refer to~\cite{Metz:2012ct} where the fragmentation contribution to $A_N$ for $p^{\uparrow} p \to h \, X$ has been computed.
Here we only write down the final expression 
\ben
P_h^0 \,\frac{ d\sigma_{UT}} {d^3\vec{P}_h} & = & - \frac{8\alpha_{\rm em}^2} {S} \,
\varepsilon_{\perp\mu\nu} \, S_{P\perp}^{\mu} \, P_{h\perp}^{\nu} \, 
\sum_q e_q^2 \int_{z_{\rm min}}^1 \frac{dz} {z^3}\,\frac{1} {S+T/z}\,\frac{1} {x}
\nonumber \\
& & \hspace{0.5cm} \times \Bigg\{\!\!-\!\frac{\pi M} {\hat{u}}\,D_1^{h/q}(z) \bigg(F_{FT}^q(x,x)-x\frac{dF_{FT}^q(x,x)} {dx}\bigg)\!\bigg[\frac{\hat{s}(\hat{s}^2+\hat{u}^2)} {2\hat{t}^{\hspace{0.025cm}3}}\bigg]
\nonumber \\
& & \hspace{1.0cm}+\,\frac{M_h} {-x\hat{u}-\hat{t}}\,\,h_{1}^{q}(x)\,\Bigg\{\!\!\bigg(\hat{H}^{h/q}(z)-z\frac{d\hat{H}^{h/q}(z)} {dz}\bigg)\!\bigg[\frac{(1-x)\hat{s}\hat{u}} {\hat{t}^{\hspace{0.025cm}2}}\bigg] 
\nonumber \\
& & \hspace{-1.cm} +\, \frac{1} {z} \, H^{h/q}(z) \bigg[ \frac{\hat{s} (\hat{s}^2 +(x-1)\hat{u}^2)} {\hat{t}^{\hspace{0.025cm}3}}\bigg] 
+ 2 z^2 \! \int_z^\infty \! \frac{dz_1} {z_1^2} \, \frac{1} {\frac{1} {z} -\frac{1} {z_{1}}} \, \hat{H}_{FU}^{h/q,\Im}(z,z_{1}) 
\bigg[ \frac{x\hat{s}^2\hat{u}} {\xi_{\hspace{0.025cm}z} \, \hat{t}^{\hspace{0.025cm}3}} \bigg]\!\Bigg\}\!\Bigg\} \,,
\label{e:lNhXUT}
\een
where we use the convention $\varepsilon^{12}_{\perp} \equiv \varepsilon^{-+12} = 1$, and $\xi_{\,z} = z/z_g$ with $1/z_{g}=(1/z-1/z_1)$. 
For the electromagnetic interaction we used both Feynman gauge and a light-cone gauge.
In either case we obtained identical results which can be considered a non-trivial cross check of the calculation.
At the operator level and in a parton model analysis, the QS function $F_{FT}^q$~\cite{Qiu:1991pp,Qiu:1991wg} can be related to the first $k_\perp$ moment of the Sivers function $f_{1T}^{\perp q}$~\cite{Boer:2003cm,Kang:2011hk},
\be
\pi \, F_{FT}^q(x,x) = \int d^2 \vec{k}_{\perp} \, \frac{\vec{k}_{\perp}^{\,2}}{2 M^2} \, f_{1T}^{\perp q}(x,\vec{k}_{\perp}^{\,2}) \Big|_{\rm SIDIS} \,,
\label{e:ETQS}
\ee
where the subscript ``SIDIS'' indicates the Sivers function probed in semi-inclusive deep-inelastic scattering. 
More information about the relation between the QS function and the Sivers effect when taking into account evolution can be found in~\cite{Kang:2011mr,Aybat:2011ge}.
The function $\hat{H}^{h/q}$ has the following relation to the Collins function $H_1^{\perp h/q}$~\cite{Yuan:2009dw,Kang:2010zzb,Metz:2012ct},
\be
\hat{H}^{h/q}(z) = z^2 \int d^2 \vec{p}_{\perp} \, \frac{\vec{p}_{\perp}^{\,2}}{2 M_h^2} \, H_1^{\perp h/q}(z,z^2\vec{p}_{\perp}^{\,2}) \,.
\label{e:HMOM}
\ee
Our definitions for both $f_{1T}^{\perp q}$ and $H_1^{\perp h/q}$ follow the so-called Trento convention \cite{Bacchetta:2004jz}. 
On the fragmentation side $\sigma_{UT}$ contains two additional twist-3 terms.
Those depend on the two-parton correlator $H^{h/q}$ and the (imaginary part of the) 3-parton correlator $\hat{H}_{FU}^{h/q}$. The underlying dynamics for these functions may be similar to the one for the Collins effect, and it turns out in fact that $\hat{H}^{h/q}, H^{h/q}$, and $\hat{H}_{FU}^{h/q,\Im}$ are not independent of each other but satisfy the relation~\cite{Metz:2012ct}
\be
H^{h/q}(z) = -2 z  \hat H^{h/q}(z)+ 2 z^3 \! \int_z^\infty \! \frac{dz_1} {z_1^2} \, \frac{1} {\frac{1} {z} -\frac{1} {z_{1}}} \, \hat{H}_{FU}^{h/q,\Im}(z,z_{1}) \,.
\label{e:relation}
\ee

Since both the Sivers function $f_{1T}^{\perp q}$ and the Collins function $H_1^{\perp h/q}$ have been extracted from experimental data~\cite{Efremov:2004tp,Anselmino:2005ea,Vogelsang:2005cs,Collins:2005ie,Efremov:2006qm,Anselmino:2007fs,Arnold:2008ap,Anselmino:2008sga,Anselmino:2013vqa,Sun:2013hua,Echevarria:2014xaa}, one has information for the twist-3 correlators $F_{FT}^q(x,x) $ and $\hat{H}^{h/q}(z)$ through Eqs.~\eqref{e:ETQS} and \eqref{e:HMOM}. 
In order to estimate the contributions from the different terms in Eq.~\eqref{e:lNhXUT}, the only unknown piece is the 3-parton correlator $\hat{H}_{FU}^{h/q,\Im}$ after taking advantage of the relation in Eq.~\eqref{e:relation}. 
In Ref.~\cite{Kanazawa:2014dca} it was argued that the fragmentation functions $H^{h/q}$ and $\hat{H}_{FU}^{h/q,\Im}$ could be the main source of the left-right asymmetry $A_N$ for $p^\uparrow \,p\to \pi\,X$. 
In our numerical estimates in the next section we will use the fitted parametrization for $\hat{H}_{FU}^{h/q,\Im}$ from~\cite{Kanazawa:2014dca}.
In principle through the process under consideration one can learn more about the fragmentation functions entering twist-3 calculations, which in turn could help one to better understand $A_N$ in proton-proton collisions where the same functions show up. 
In practice, however, this may be difficult due to potentially large NLO radiative corrections for $\ell \, N\to h \, X$~\cite{Kang:2011jw,Kang:2013lga}. 
Below we will return to this point.

%
% 3. Section: Numerical estimates
% ===========================
%
\section{Numerical estimates \label{sectionIII}}
\noindent 
In this section, we will estimate $A_N$ based on the LO formulas in Eqs.~\eqref{e:lNhXUU} and \eqref{e:lNhXUT}.  
We will study in detail the contributions from the soft-gluon pole term involving $F_{FT}^q$, and the fragmentation term involving  $\hat{H}^{h/q}$, ${H}^{h/q}$, and $\hat{H}^{h/q,\Im}_{FU}$.  
Throughout we use the GRV98 unpolarized parton distributions \cite{Gluck:1998xa} and the DSS unpolarized fragmentation functions \cite{deFlorian:2007aj}.  
We calculate the QS function $F_{FT}^q$ using Eq.~\eqref{e:ETQS} and the Sivers function of Ref.~\cite{Anselmino:2008sga} extracted from SIDIS data. 
The twist-3 fragmentation function $\hat{H}^{h/q}$ is calculated by using Eq.~\eqref{e:HMOM} and the Collins function extracted from SIDIS and $e^+e^-$ data in Ref.~\cite{Anselmino:2013vqa}.  
The transversity function $h_1^q$ is also taken from Ref.~\cite{Anselmino:2013vqa}.  
Note that antiquark transversity functions are neglected throughout since no information exists on their extraction. 
The function $\hat{H}^{h/q,\Im}_{FU}$ was fitted to data on $A_N$ in {\it pp} scattering in Ref.~\cite{Kanazawa:2014dca}, and we will use its parametrization, while $H^{h/q}$ is fixed through Eq.~\eqref{e:relation}. 
For simplicity we assume that the twist-3 correlators follow the same DGLAP scale dependence as the twist-2 counterparts $f_1^q$ or $D_1^{h/q}$.
For more information on the proper evolution of 3-parton correlators we refer to~\cite{Kang:2008ey,Zhou:2008mz,Vogelsang:2009pj,Braun:2009mi,Kang:2010xv,Schafer:2012ra,Ma:2012xn,Kang:2012em,Ma:2012ye,Schafer:2013opa}.

% ===========================
\subsection{Single spin asymmetry}
In order to compare our calculation to the spin asymmetry measured by the HERMES Collaboration~\cite{Airapetian:2013bim} we first need to carefully consider the conventions.
In Ref.~\cite{Airapetian:2013bim} the transverse SSA is denoted by $A_{UT}^{\sin{\Psi}}$ and defined through
\ben
d\sigma = d\sigma_{UU} ( 1 + S_{P\perp} A_{UT}^{\sin{\Psi}} \sin{\Psi} ) \,.
\een
Here the $\sin{\Psi}$ azimuthal dependence is determined from the vector product $\vec S_{P\perp} \cdot (\vec P_h \times \vec l\,)$ where, as stated above, $\vec S_{P\perp}$ is the (transverse) spin vector of the target, and $\vec l$ and $\vec P_h$ are the three-momenta of the incident lepton and of the final-state hadron, respectively. 
The asymmetry is defined in the lepton-nucleon center-of-mass frame such that the lepton moves in the $+z$ direction, while the transversely polarized nucleon moves along the $-z$ direction. 
While in the {\it pp} case the transversely polarized nucleon typically defines the $+z$ direction, it is important to realize that the definition of $A_{UT}^{\sin{\Psi}}$ fully agrees with the one for $A_N$ used for {\it pp} collisions~\cite{Adams:1991cs,Adler:2005in,Adare:2013ekj,Adams:2003fx,Abelev:2008qb,Adamczyk:2012xd,Arsene:2008mi,Bland:2013pkt}.
Note also that in the HERMES convention positive Feynman $x$ (which we denote by $x_F^H$) corresponds to hadrons going in the direction of the lepton or backwards with respect to the target nucleon. 
This convention has the opposite sign compared to $x_F$ used in the {\it pp} case~\cite{Adams:1991cs,Adler:2005in,Adare:2013ekj,Adams:2003fx,Abelev:2008qb,Adamczyk:2012xd,Arsene:2008mi,Bland:2013pkt}, i.e., $x_F^H = -x_F$. 
With the coordinate system specified in Section II and the spin vector of the nucleon pointing in the $+y$ direction we have
\ben
\epsilon_{\perp\mu\nu} S_\perp^\mu\Php^\nu = -\Php \, , \qquad
x_F \equiv \frac{2 P_{hz}}{\sqrt{S}} = -x_F^{\rm H} \,,
\een
 where $\Php \equiv |\vec{P}_{h\perp}|$. 
The differential cross section can then be written as
\ben
P_h^0\, \frac{d\sigma_{UT}}{d^3 P_h} = \sqrt{4 \frac{\Php^2}{S}+x_F^2}\, \frac{d \sigma_{UT}}{d x_F d^2 \Php} \,,
\een
and the spin asymmetry $A_N$ is given by
\ben
A_N (x_F,\Php)  \equiv   \frac{\sqrt{4 \frac{\Php^2}{S}+x_F^2}\, \frac{d \sigma_{UT}}{d x_F d^2 \Php}}{\sqrt{4 \frac{\Php^2}{S}+x_F^2}\, \frac{d \sigma_{UU}}{d x_F d^2 \Php}} 
= A_{UT}^{\sin{\Psi}}(-x_F^H,\Php) \,.
\label{e:AnAUT}
\een
 
The Mandelstam variables $T$ and $U$ can be expressed in terms of $x_F$ and $\Php$ as
\ben
T = -\sqrt{S}  \sqrt{ \Php^2 + x_F^2  \frac{S}{4} }  + x_F  \frac{S}{2} \,,
\qquad
U = -\sqrt{S}  \sqrt{ \Php^2 + x_F^2  \frac{S}{4} }  - x_F  \frac{S}{2} \,. 
\label{e:kin}
\een
These relations will help us to better understand the kinematical regions that are covered in the integration in Eqs.~\eqref{e:lNhXUU} and \eqref{e:lNhXUT}.
In particular, let us consider a situation $S \gg \Php^2$ and $ x_F \rightarrow -1$ ($ x_F^{\rm H} \rightarrow 1$). 
It is easy to see that in this case $T\rightarrow -S$, $U\rightarrow 0$.
If $ x_F \rightarrow 1$ ($ x_F^{\rm H} \rightarrow -1$),  we have $T\rightarrow 0$, $U\rightarrow -S$.  
We may conclude from Eq.~\eqref{e:xdef} that for $ x_F \rightarrow 1$ ($ x_F^{\rm H} \rightarrow -1$) the region of $x$ will be concentrated around $1$, i.e., the large-$x$ region, and for $ x_F \rightarrow -1$ ($ x_F^{\rm H} \rightarrow 1$) $x$ will be in the region $[0,1]$, i.e., relatively small-$x$ region. 
On the other hand, the region explored in $z$ spans from $z_{\rm min}= -(T+U)/S$ to $1$ and is obviously symmetric with respect to $x_F \leftrightarrow -x_F$. 
The region of $z$ will shrink to $1$ for both $x_F \rightarrow \pm 1$. 

\begin{figure*}[hbt]
\centering
      \includegraphics[scale=0.6]{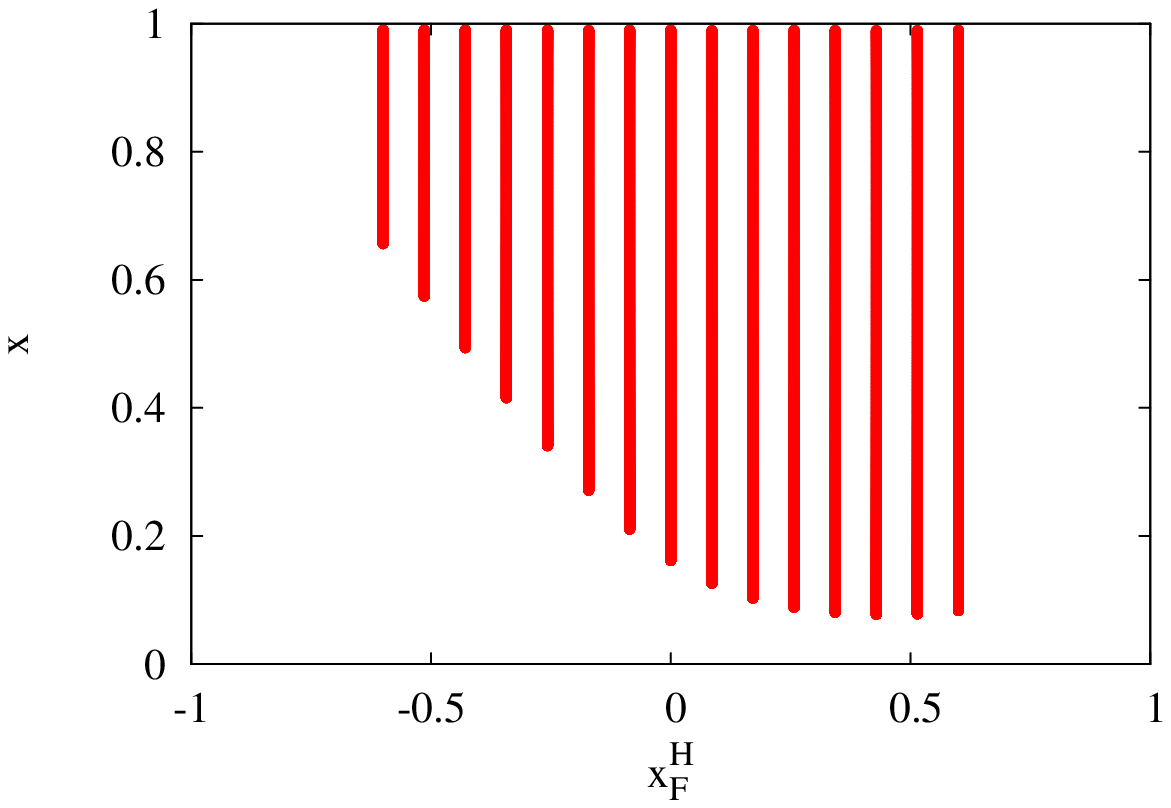}
      \includegraphics[scale=0.6]{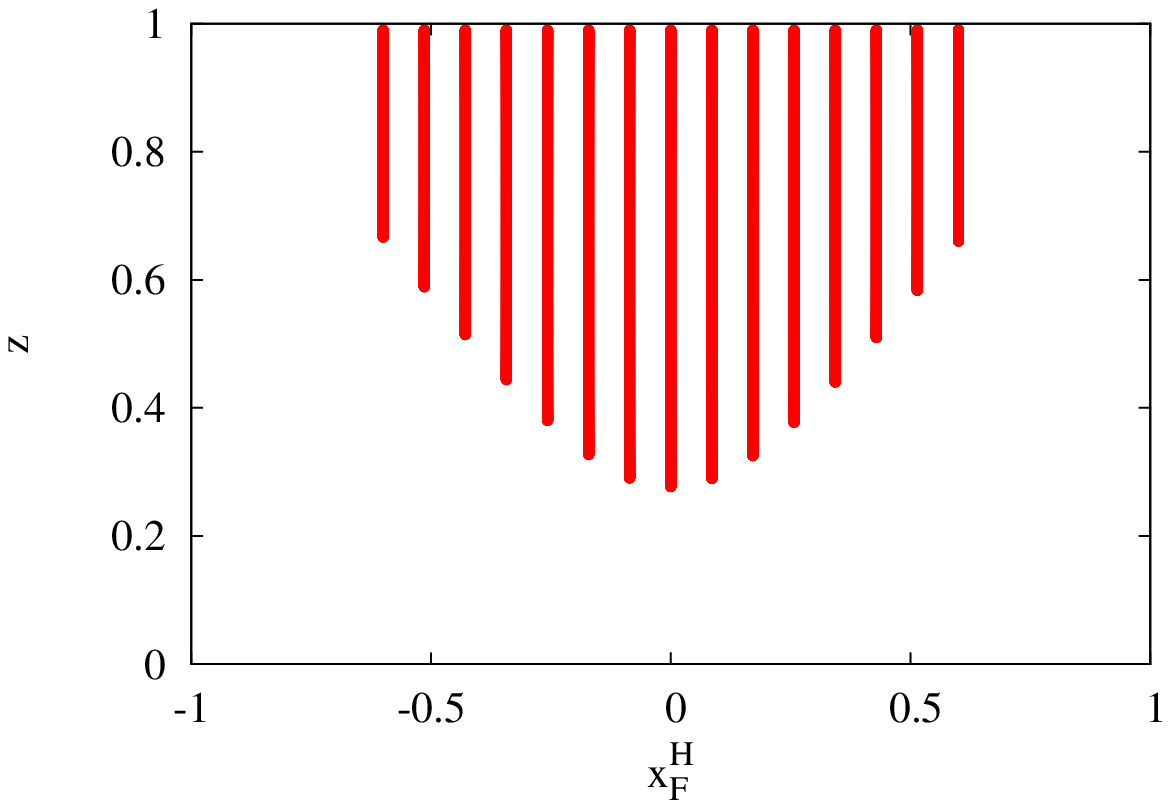}
\caption{Region covered in $x$ (left panel) and in $z$ (right panel) as a function of $x_F^{\rm H}$.}
\label{fig:kin}
\end{figure*}

It is good to discuss the uncertainties in our formalism, which mainly come from $F_{FT}^q$, $h_{1}^{q}$, $\hat{H}^{h/q}$, and $\hat{H}^{h/q,\Im}_{FU}$. The errors of $F_{FT}^q$, $h_{1}^{q}$ and $\hat{H}^{h/q}$ are propagated from the errors of TMD distributions which were extracted in \cite{Anselmino:2008sga, Anselmino:2013vqa}. The errors of these TMD distributions were estimated in \cite{Anselmino:2008sga,Anselmino:2013vqa} using the method described in Appendix A of \cite{Anselmino:2008sga}. Let us briefly recall this method: In order to calculate the errors we generate randomly 200 sets of parameters 
for each of the distributions considered such that each of the set gives $\chi^2$ which is within $\Delta \chi^2$ (chosen to correspond to 95\% of confidence level) above the minimum $\chi^2_{\rm min}$ reached by global minimization on the corresponding experimental data set. In order to calculate the errors on the observables, we then compute these observables using all 200 sets and find a minimum and maximum for each point. By doing so we plot an error band for all curves in this paper: in other words, the uncertainty band on all the plots in the rest of the paper contains only the errors of these TMD distributions. 
Notice that through this procedure we have a joint estimation 
of errors for transversity and Collins FF simultaneously as they 
enter into observables together. It is natural to expect that uncertainties 
of corresponding functions grow in the region where the experimental data are 
not available,
for example for the Sivers function and transversity at large $x$ (or negative $x_F^{H}$). From this simple analysis, we can conclude that in the region of $x_F \rightarrow 1$ ($ x_F^{H} \rightarrow -1$)  our calculations will have the largest uncertainties  based on the uncertainties in the TMD functions. This is because this $x$-range probes the as yet unexplored regions in SIDIS of large $x, z\rightarrow 1$. On the other hand in the regime $x_F \rightarrow -1$ ($ x_F^{H} \rightarrow 1$) we expect to have smaller uncertainties based on the uncertainties in the TMD functions as far as this region of $x$ corresponds to the kinematical regime already explored in SIDIS. In order to corroborate these findings we also present the numerical computation of $x$ and $z$ as a function of $x_F^H$ in the case of HERMES kinematics at $\sqrt{S} = 7.25$ GeV for $\Php=1$ GeV in Fig.~\ref{fig:kin}. One can see that, in particular, in the region of positive $ x_F^{H}$ (negative $ x_F$),  the region of $x$ indeed corresponds to the region explored by SIDIS data.  Note in the cross section for $\ell p$ that $z$ is integrated over while $x$ is fixed once $z$ is known (or vice versa).  But in the case of $pp$ collisions, once $z$ is known, only a minimum $x$-value $x_{\rm min}$ is fixed, with an $x$-integration evaluated from $x_{\rm min}$ to 1.

Nevertheless, one must keep in mind that the function $\hat{H}^{h/q,\Im}_{FU}$ was fitted to experimental data in proton-proton scattering which are in the large positive $x_F$ range (i.e., large negative $x_F^{H}$ region) not explored by inclusive hadron production in lepton-proton scattering at HERMES. Error bands for these functions were not computed in Ref.~\cite{Kanazawa:2014dca}.  However, one might speculate in the region of $x_F^{H} > 0$ ($x_F < 0$), where limited information from $pp$ collisions exists and none is available for charged pions, there are large errors.  Without the uncertainty of $\hat{H}^{h/q,\Im}_{FU}$ included, the error bands in the plots are thus underestimated in this $x_F^H > 0$ region.  In addition, even in the $x_F > 0$ region covered by the $pp$ data, one has uncertainties in $\hat{H}^{h/q,\Im}_{FU}$ due to uncertainties in the Sivers, Collins, and transversity functions that were used as inputs in the analysis of Ref.~\cite{Kanazawa:2014dca}.  This is readily seen in the different fits obtained in Ref.~\cite{Kanazawa:2014dca} when using two different extractions of the Sivers functions.  Since $z_{\rm min}$ increases as $x_F$ increases, this implies large uncertainties in $\hat{H}^{h/q,\Im}_{FU}$ in the large-$z$ region covered by the HERMES data (see Fig.~\ref{fig:kin}).  There are also uncertainties in $\hat{H}^{h/q,\Im}_{FU}$ related to the neglect of soft-fermion poles in the $pp$ reaction, which may play some role in $A_N$ for that process~\cite{Koike:2009ge}.  In any case, a complete analysis of both $\ell p$ and $pp$ asymmetry data within the factorization formalisms (with enough accuracy in theoretical calculations) should better constrain $\hat{H}^{h/q,\Im}_{FU}$, and in turn help to more thoroughly understand this fragmentation mechanism that could underlie single-spin asymmetries in proton-proton collisions \cite{Kanazawa:2014dca}.

%%%%%%%%%%%%%%%%%%%%%%%%%%%%%%%%%%%
%%%%%%%%%%%%%%%%%%%%%%%%%%%%%%%%%%%

% ===========================
\subsection{Comparison with the experimental data}
 
In the following we will plot $A_N (-x_F,\Php) = A_{UT}^{\sin{\Psi}}(x_F^H, \Php)$ as a function of $x_F^H$ and $\Php$.
It is important to realize that for the process at hand, $\ell\,N\to h\,X$, only the hadron transverse momentum $\Php$ can serve as the hard scale. 
We thus choose the renormalization scale for both parton distributions and fragmentation functions as $\Php$, which has to satisfy $\Php\gg \Lambda_{\rm QCD}$ to ensure the use of collinear factorization formalism. 
With this in mind, we therefore only compare with the HERMES data in Ref.~\cite{Airapetian:2013bim} with $\Php \geq 1$ GeV.  
We note that almost all of this data is from quasi-real photoproduction (i.e., $Q^2 \sim 0 \; \rm{GeV}^2$).  
We will address later how this could affect the comparison between theory and experiment.

%%%%%%%%%%%%%%%%%%%%%%%%%%%%%%%%%%%
\begin{figure*}[hbt]
\centering
      \includegraphics[scale=0.425]{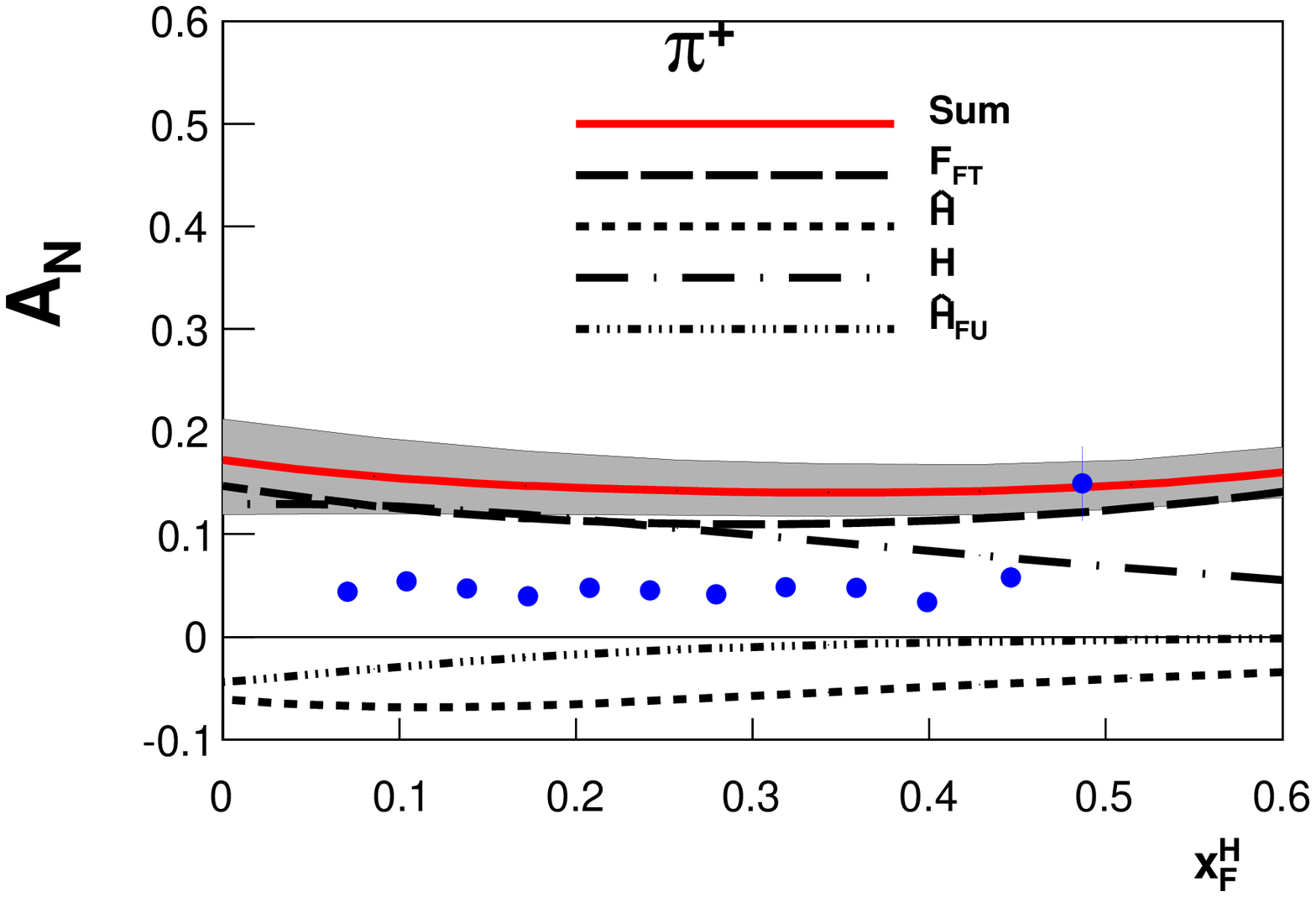}\hspace{-0.7cm}
      \includegraphics[scale=0.425]{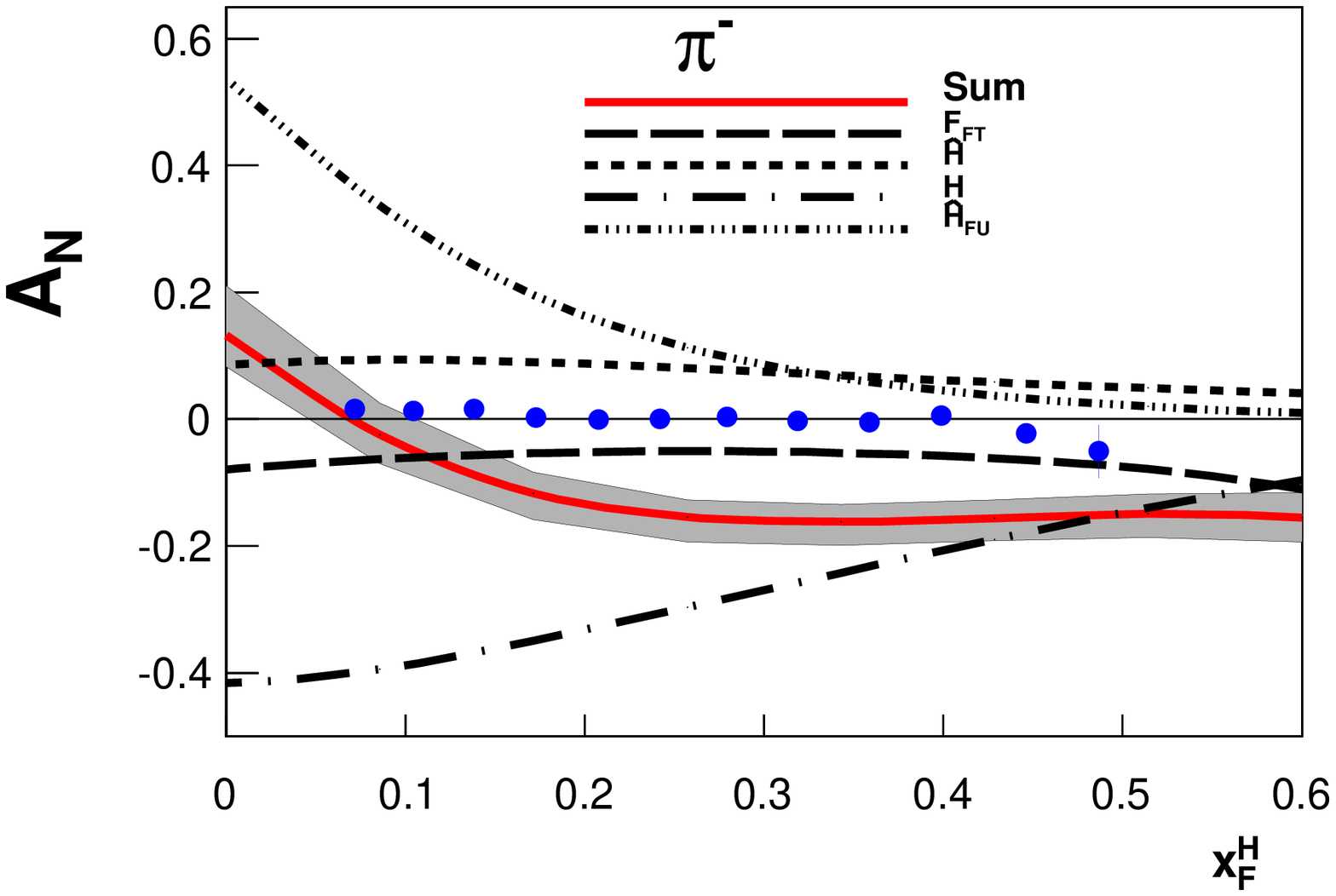}
\caption{$A_N$ as a function of $x_F^H$ for $\pi^+$ (left panel) and $\pi^-$ (right panel) production at $\Php = 1 \, \rm{GeV}$ for lepton-proton collisions at $\sqrt{S} = 7.25 \, \rm{GeV}$. 
The data are from Ref.~\cite{Airapetian:2013bim}. 
The solid line corresponds to the sum of all contributions. 
The $F_{FT}^q$ contribution is the dashed line, the $\hat{H}^{h/q}$ contribution is the dotted line, the ${H}^{h/q} $contribution is the dot-dashed line, and the $\hat{H}^{h/q}_{FU}$ contribution is the 3-dotted-dashed line. 
The error band comes from uncertainties in the Sivers, Collins, and transversity functions estimated in Refs.~\cite{Anselmino:2008sga,Anselmino:2013vqa}.  
Note that positive $x_F^{\rm H}$ corresponds to pions in the backward direction with respect to the target proton.}
\label{fig:an_hermes_pipm_xf}
\end{figure*}
%%%%%%%%%%%%%%%%%%%%%%%%%%%%%%%%%%%

In Fig.~\ref{fig:an_hermes_pipm_xf} we plot $A_N$ as a function of $x_F^{\rm H}$ for $\pi^+$ and $\pi^-$ production with $1 < \Php < 2.2 \; \rm{GeV}$ ($\langle \Php\rangle \simeq 1$ GeV) for lepton-proton collisions at HERMES energy $\sqrt{S} = 7.25$ GeV \cite{Airapetian:2013bim}. 
For $\pi^+$ the contribution coming from $F_{FT}^q$ related to the Sivers effect is positive for all $x_F$.  
The contribution from $\hat{H}^{h/q}$ is of opposite sign and smaller in absolute value than that from $F_{FT}^q$.  
The contribution from ${H}^{h/q}$ is positive and that from $\hat{H}^{h/q,\Im}_{FU}$ is negative, and their sum is similar in absolute value to the contribution from $\hat{H}^{h/q}$.  
In fact those three contributions almost cancel each other leaving a nearly vanishing fragmentation piece. 
The resulting asymmetry is close to the contribution from  $F_{FT}^q$ and is larger than the experimental data, as clearly seen in the figure. 
The experimental data are around 5\% and our computations result in a positive asymmetry of about 15\%.

On the other hand, for $\pi^-$ the contribution coming from $F_{FT}^q$ is negative for positive  $x_F^{\rm H}$
and the contribution from $\hat{H}^{h/q}$ is of opposite sign and comparable to that from $F_{FT}^q$.  The contribution for positive $x_F^H$ from ${H}^{h/q}$ is negative and from $\hat{H}^{h/q,\Im}_{FU}$ is positive.
The fragmentation piece contributes roughly the same as $F_{FT}^q$ does at moderate $x_F^H$ but begins to dominate at smaller (and negative) $x_F^H$. 
Our computations result in a negative asymmetry of about $-$15\% in the positive $x_F^{\rm H}$ region whereas the experimental data are close to zero.
 
%%%%%%%%%%%%%%%%%%%%%%%%%%%%%%%%%%%
\begin{figure*}[hbt]
\centering
      \includegraphics[scale=0.425]{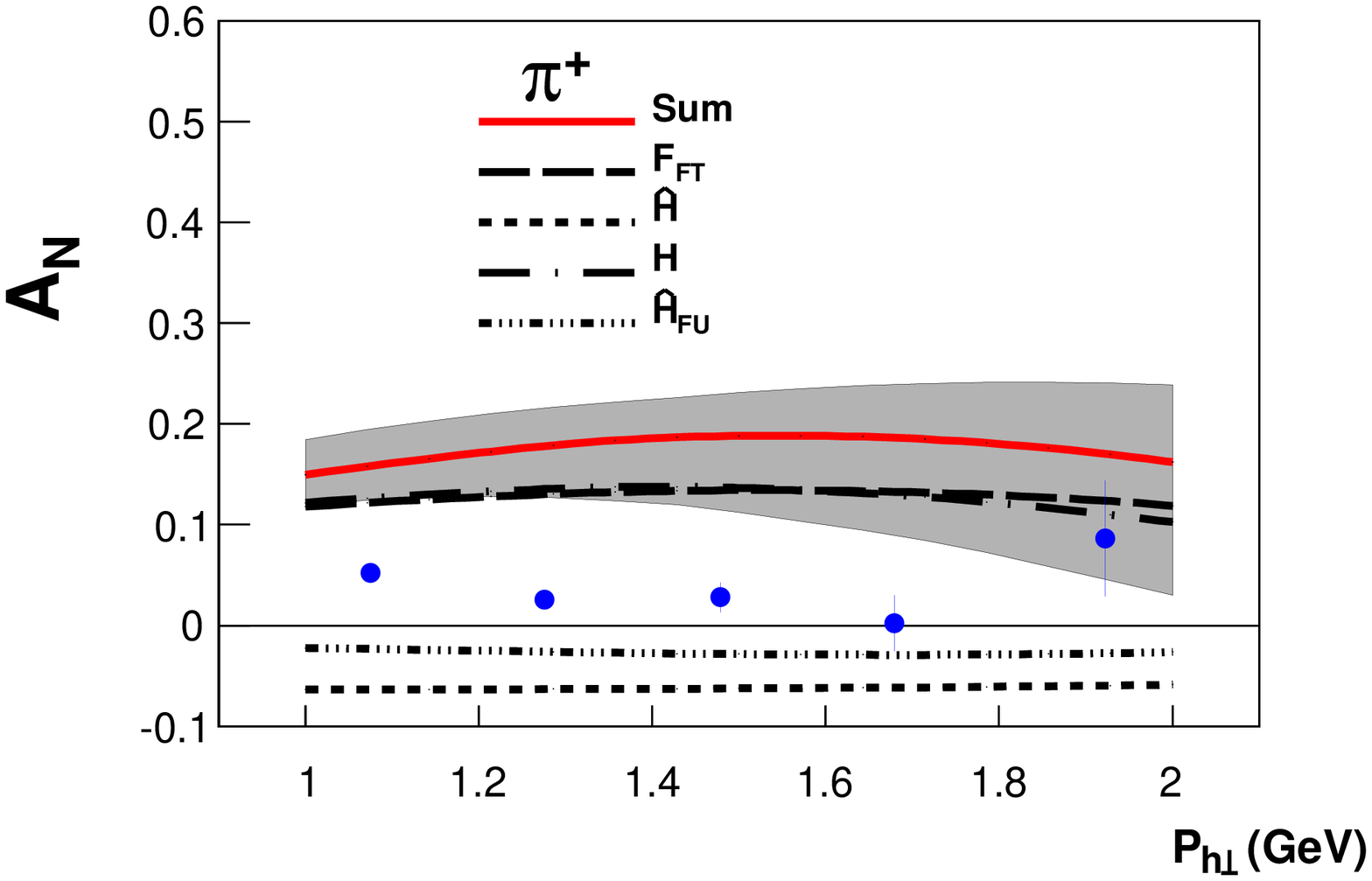}\hspace{-0.7cm}
      \includegraphics[scale=0.425]{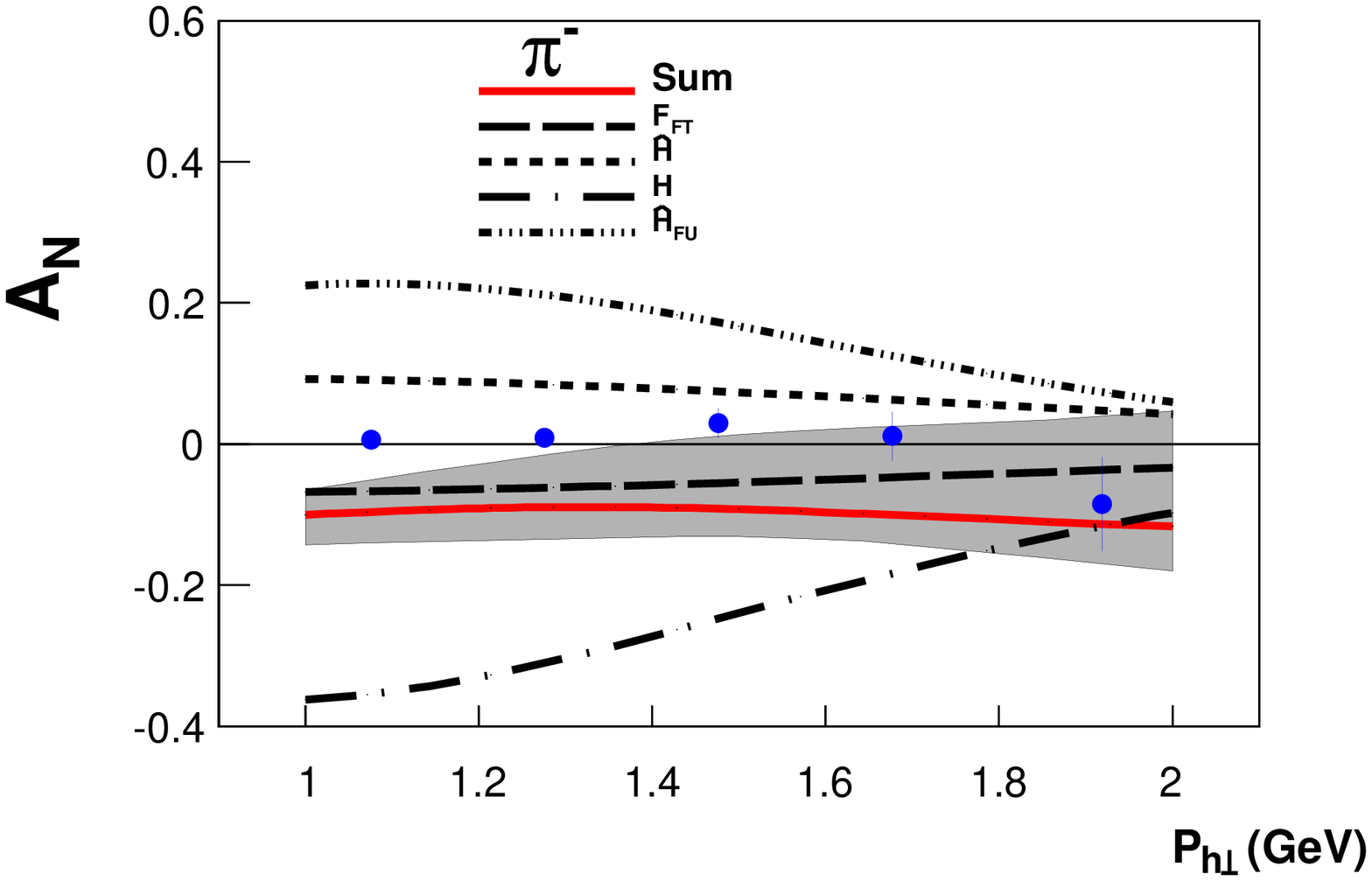}
\caption{$A_N$ as function of $\Php$ for $\pi^+$ (left panel) and $\pi^-$ (right panel) production for lepton-proton collisions at
$0.1 < x_F^{\rm H} < 0.2$  ($\langle x_F^{\rm H} \rangle \simeq 0.15$) and $\sqrt{S} = 7.25 \; \rm{GeV}$. 
The data are from Ref.~\cite{Airapetian:2013bim}.  
The description of lines is the same as in Fig~\ref{fig:an_hermes_pipm_xf}.}
\label{fig:an_hermes_pipm_pt}
\end{figure*}
%%%%%%%%%%%%%%%%%%%%%%%%%%%%%%%%%%%
In Fig.~\ref{fig:an_hermes_pipm_pt} we plot $A_N$ as function of $\Php$ for lepton-proton collisions at HERMES energy $\sqrt{S} = 7.25 \; \rm{GeV}$~\cite{Airapetian:2013bim}.
The general trends for all contributions are similar to those for the $x_F$ dependence shown in Fig.~\ref{fig:an_hermes_pipm_xf} and described above. 
One may sense, though, that, as could have been expected, our LO calculation is doing better towards larger values of $P_{h\perp}$.

%%%%%%%%%%%%%%%%%%%%%%%%%%%%%%%%%%%
\begin{figure*}[hbt]
\centering
      \includegraphics[scale=0.425]{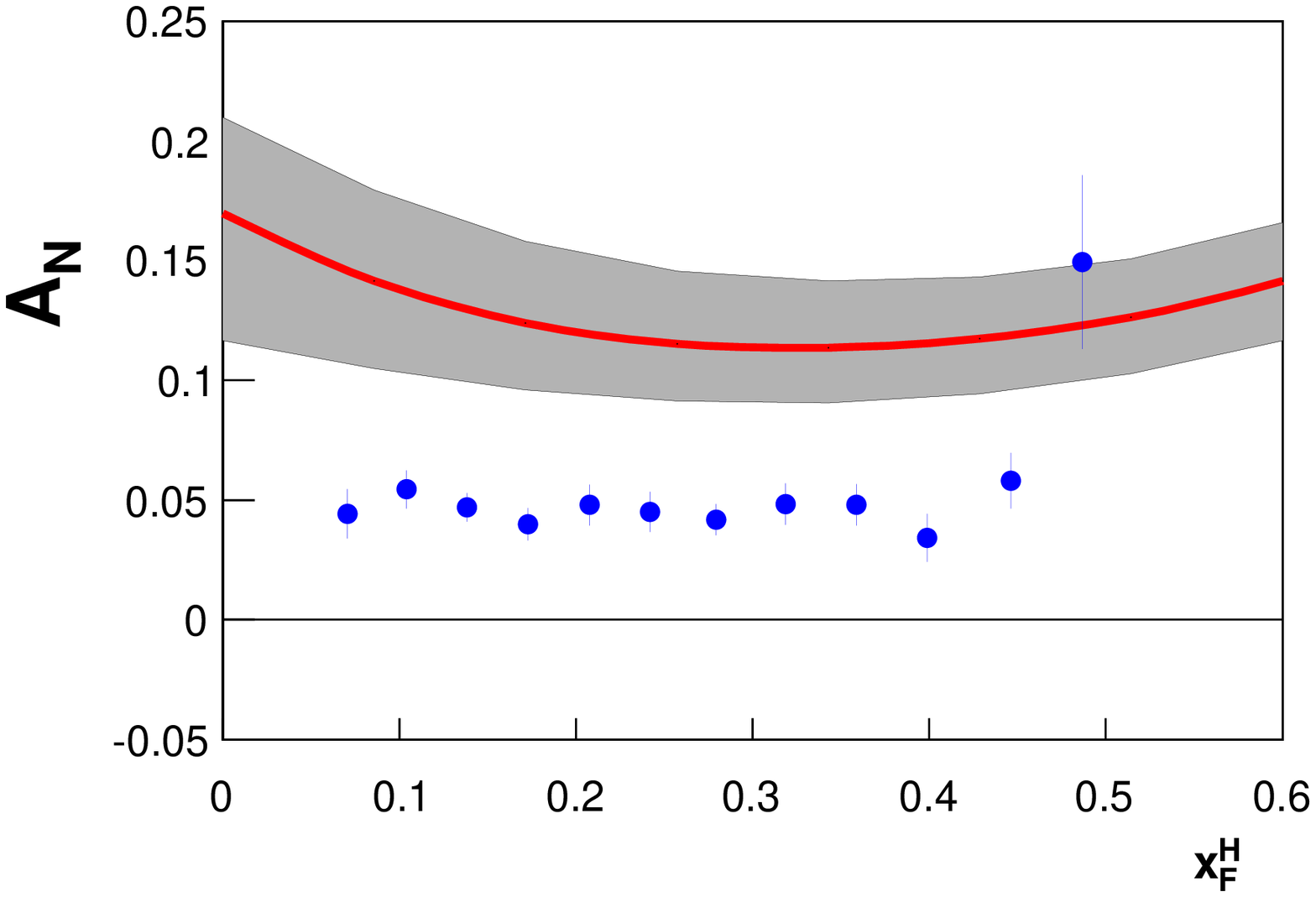}\hspace{-0.7cm}
      \includegraphics[scale=0.425]{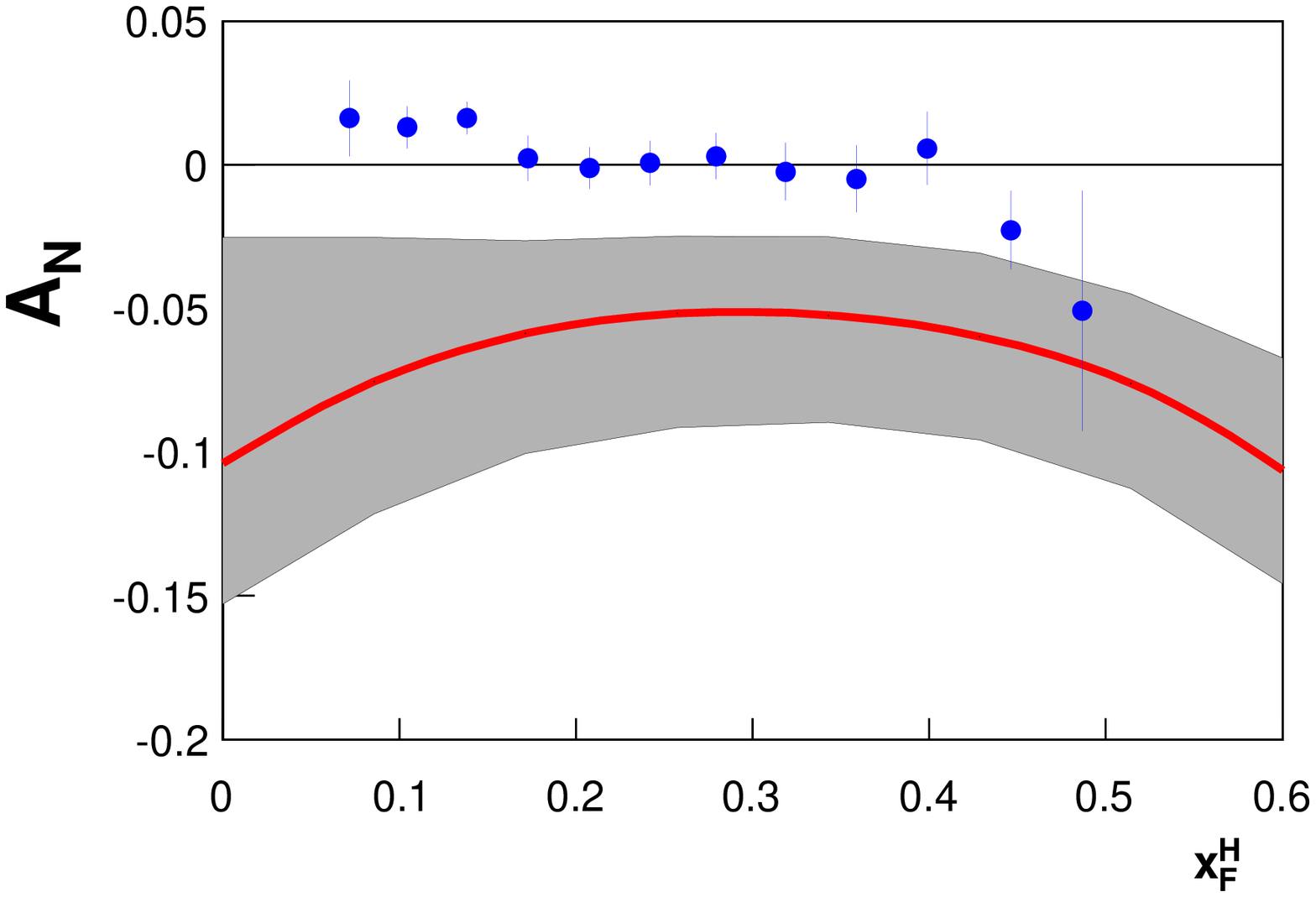}
\caption{$A_N$ as a function of $x_F^H$ for $\pi^+$ (left panel) and $\pi^-$ (right panel) production at $\Php = 1 \; \rm{GeV}$ and $\sqrt{S} = 7.25 \; \rm {GeV}$. 
The data are from  Ref.~\cite{Airapetian:2013bim}.  
The solid line corresponds to sum of all contributions with $\hat{H}^{h/q,\Im}_{FU} = 0$.}
\label{fig:an_hermes_HF_pipm_xf}
\end{figure*}
%%%%%%%%%%%%%%%%%%%%%%%%%%%%%%%%%%%

Before we proceed, let us elaborate more on the contribution due to the 3-parton correlator $\hat{H}^{h/q,\Im}_{FU}$. 
According to Ref.~\cite{Kanazawa:2014dca}, $\hat{H}^{h/q,\Im}_{FU}$, in particular through its contribution to $H^{h/q}$ via Eq.~\eqref{e:relation}, might play a critical role for the description of $A_N$ in $p^{\uparrow} p \to h \, X$ in the collinear twist-3 approach.
In Fig.~\ref{fig:an_hermes_HF_pipm_xf} we present our computations for $A_N$ when $\hat{H}^{h/q,\Im}_{FU}$ is switched off.
(Note that setting $H^{h/q}$ and $\hat{H}^{h/q,\Im}_{FU}$ to zero simultaneously would also imply $\hat{H}^{h/q} = 0$ due to the relation in Eq.~\eqref{e:relation}.) 
Comparing with Fig.~\ref{fig:an_hermes_pipm_xf} one observes that $A_N^{\pi^+}$ does not change very much.
On the other hand, the influence on $A_N^{\pi^-}$ is quite significant.
In the $x_F^H$ region of the HERMES data, the magnitude of the asymmetry is reduced by about a factor two.
The influence of switching off $\hat{H}^{h/q,\Im}_{FU}$ is most dramatic in the region of relatively small $x_F^H$ (in fact also negative $x_F^H$ as we have checked) where $A_N^{\pi^-}$ even changes sign.
In that region an EIC could provide unique information as we discuss in more detail in Section III C.

%%%%%%%%%%%%%%%%%%%%%%%%%%%%%%%%%%%
\begin{figure*}[hbt]
\centering
      \includegraphics[scale=0.425]{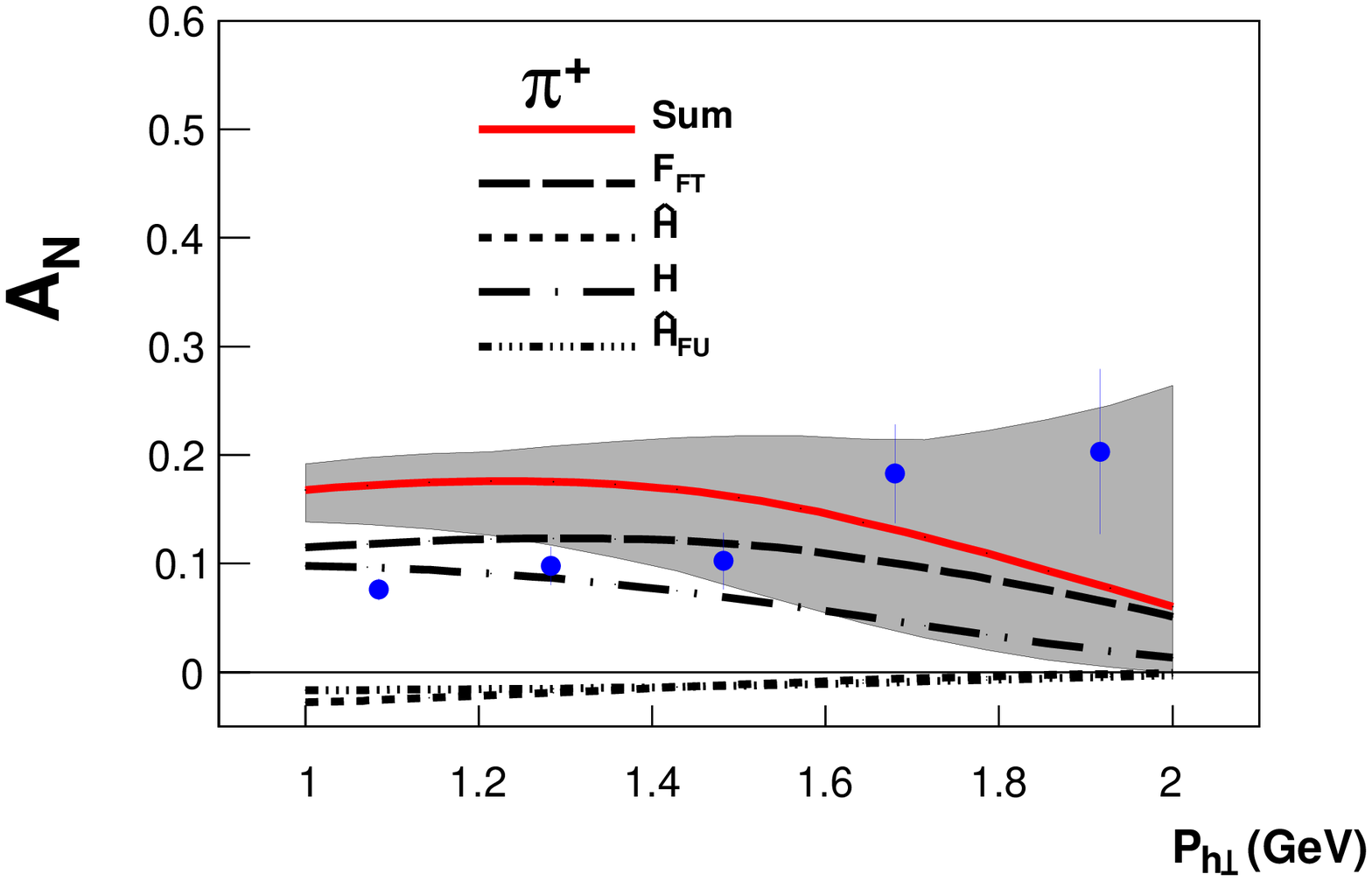}\hspace{-0.7cm}
      \includegraphics[scale=0.425]{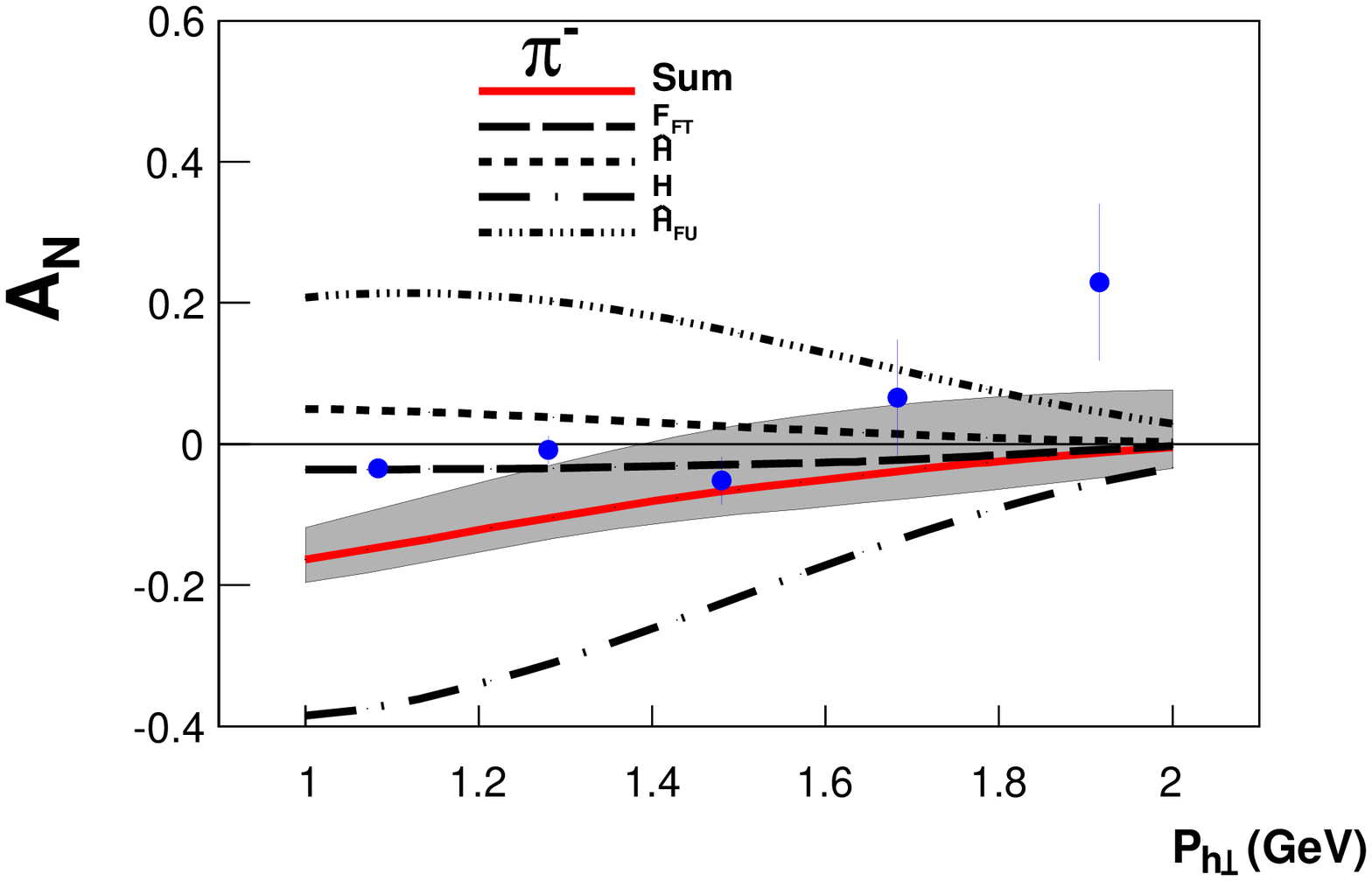}
\caption{$A_N$ as function of $\Php$ for $\pi^+$ (left panel) and $\pi^-$ (right panel) production at $\langle x_F^{\rm H} \rangle \approx 0.2$  $(0.2 < z< 0.7)$ and $\sqrt{S} = 7.25 \; \rm{GeV}$ for the ``DIS" sub-set of the data from  Ref.~\cite{Airapetian:2013bim}. 
The description of lines is the same as in Fig~\ref{fig:an_hermes_pipm_xf}.}
\label{fig:an_hermes_pipm_02z07_xf}
\end{figure*}
%%%%%%%%%%%%%%%%%%%%%%%%%%%%%%%%%%%

%%%%%%%%%%%%%%%%%%%%%%%%%%%%%%%%%%%
\begin{figure*}[hbt]
\centering
      \includegraphics[scale=0.425]{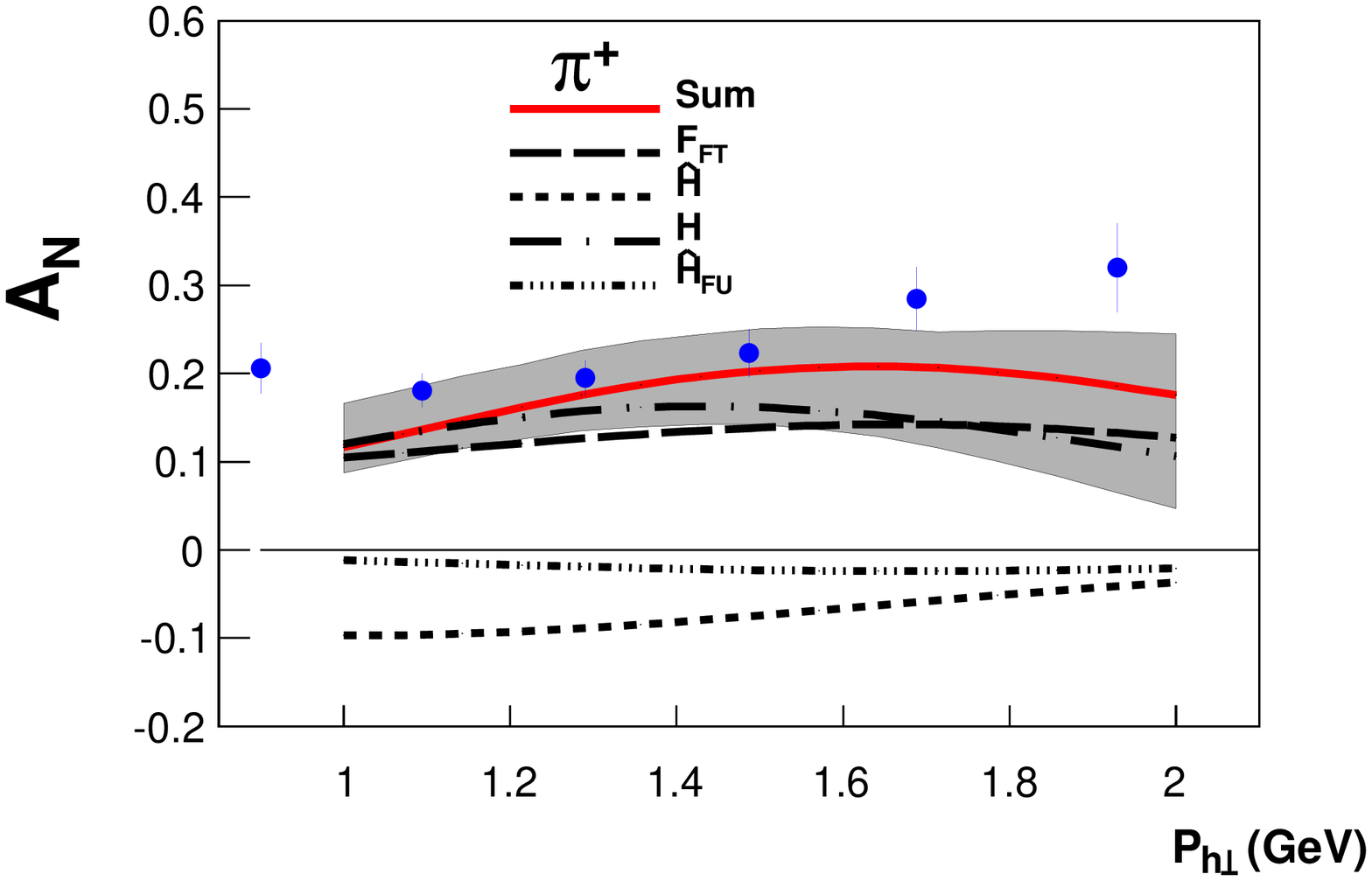}\hspace{-0.7cm}
      \includegraphics[scale=0.425]{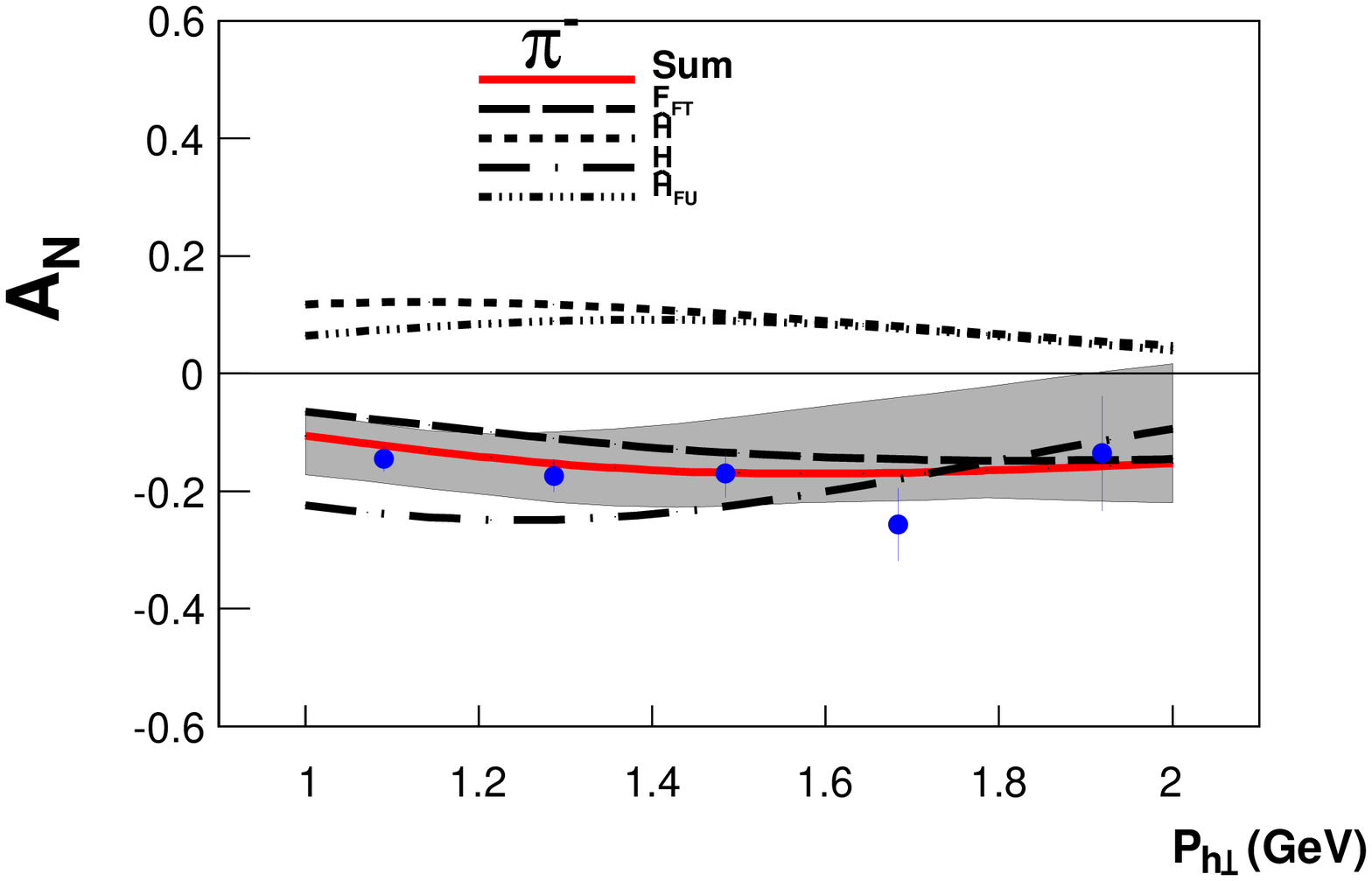}
 \caption{$A_N$ as function of $\Php$ for $\pi^+$ (left panel) and $\pi^-$ (right panel) production at $\langle x_F^{\rm H}\rangle \approx 0.27$ $(z > 0.7)$ and $\sqrt{S} = 7.25 \; \rm{GeV}$ for the ``DIS" sub-set of the data from  Ref.~\cite{Airapetian:2013bim}. The description of lines is the same as in Fig~\ref{fig:an_hermes_pipm_xf}.}
\label{fig:an_hermes_pipm_z07_xf}
\end{figure*}
%%%%%%%%%%%%%%%%%%%%%%%%%%%%%%%%%%%

HERMES also explored several sub-sets of data where the outgoing lepton was detected and photon virtuality $Q^2 > 1 \;\rm{GeV}^2$ was guaranteed, which were referred as ``DIS'' subsets~\cite{Airapetian:2013bim}. 
This subset was divided into two regions of $z$: $0.2 < z < 0.7$ ($\langle x_F^{\rm H} \rangle \approx 0.2$) and $z > 0.7$ ($\langle x_F^{\rm H} \rangle \approx 0.27$). 
Even though strictly speaking these data sets correspond to semi-inclusive rather then fully inclusive hadron production, we will nevertheless compare our calculations with these measurements.
In Fig.~\ref{fig:an_hermes_pipm_02z07_xf} we plot $A_N$ for $\pi^+$ and $\pi^-$ production as a function of $\Php$ for $0.2 < z < 0.7$. 
One can see that for  $\pi^+$ the $F_{FT}^q$ and ${H}^{h/q}$ terms dominate, while the $\hat{H}^{h/q,\Im}_{FU}$ and $\hat{H}^{h/q}$ pieces are negligible. 
For $\pi^-$ the contribution from $F_{FT}^q$ becomes smaller, and the $H^{h/q}$ and $\hat{H}^{h/q,\Im}_{FU}$ terms are sizable (with opposite sign) but decrease quickly with increasing $P_{h\perp}$. 
A different pattern emerges  for the $z > 0.7$ subset which we plot in Fig.~\ref{fig:an_hermes_pipm_z07_xf}. 
In this case, the contributions from $\hat{H}^{h/q}$ and $\hat{H}^{h/q,\Im}_{FU}$ almost cancel the contribution from $H^{h/q}$,  and the asymmetry is close to the result for the contribution from $F_{FT}^{q}$.  
Overall, the theoretical curves are much closer to the experimental data for both $\pi^+$ and $\pi^-$ production.
%%%%%%%%%%%%%%%%%%%%%%%%%%%%%%%%%%%
\begin{figure*}[hbt]
\centering
      \includegraphics[scale=0.425]{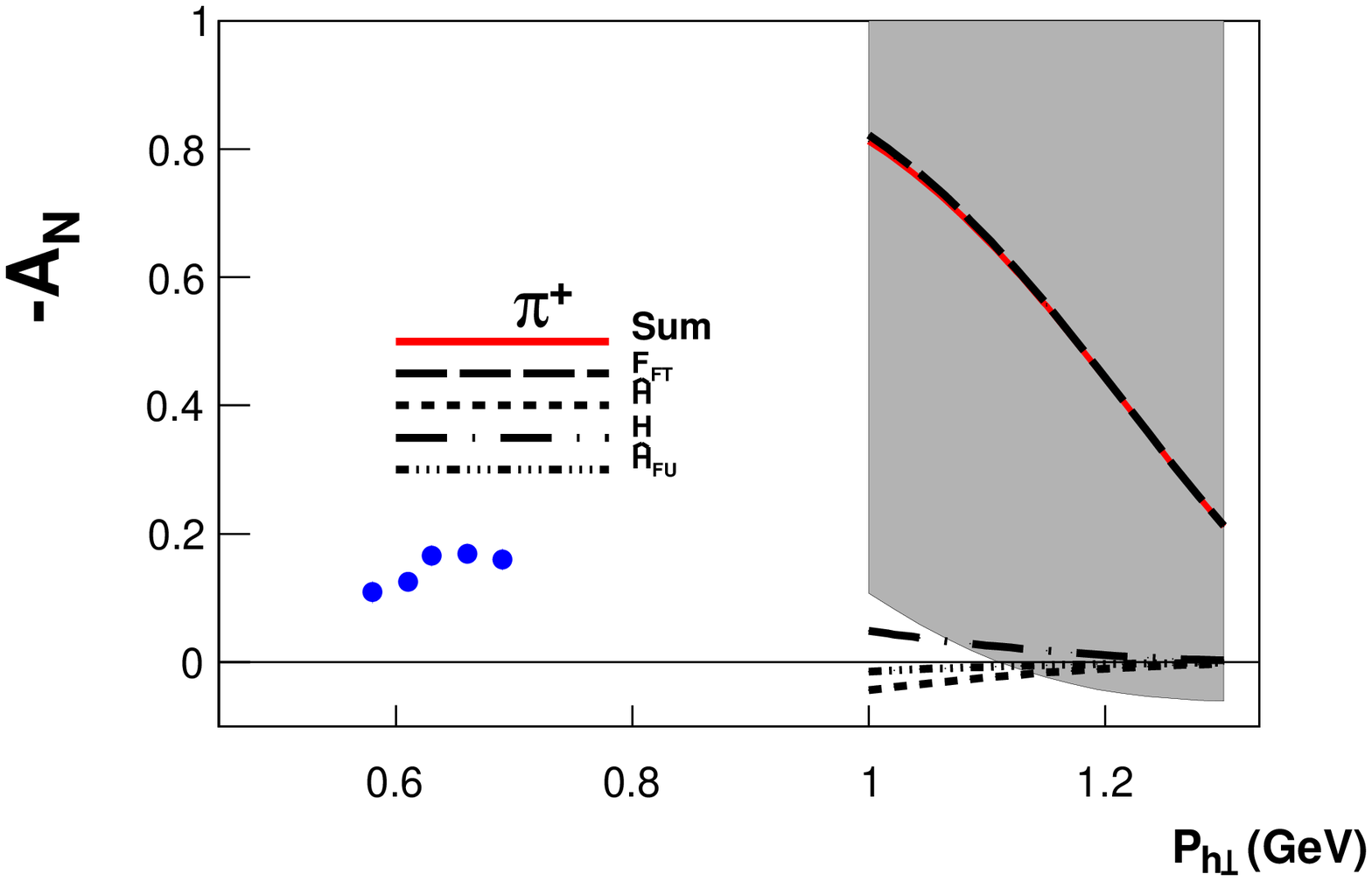}\hspace{-0.7cm}
      \includegraphics[scale=0.425]{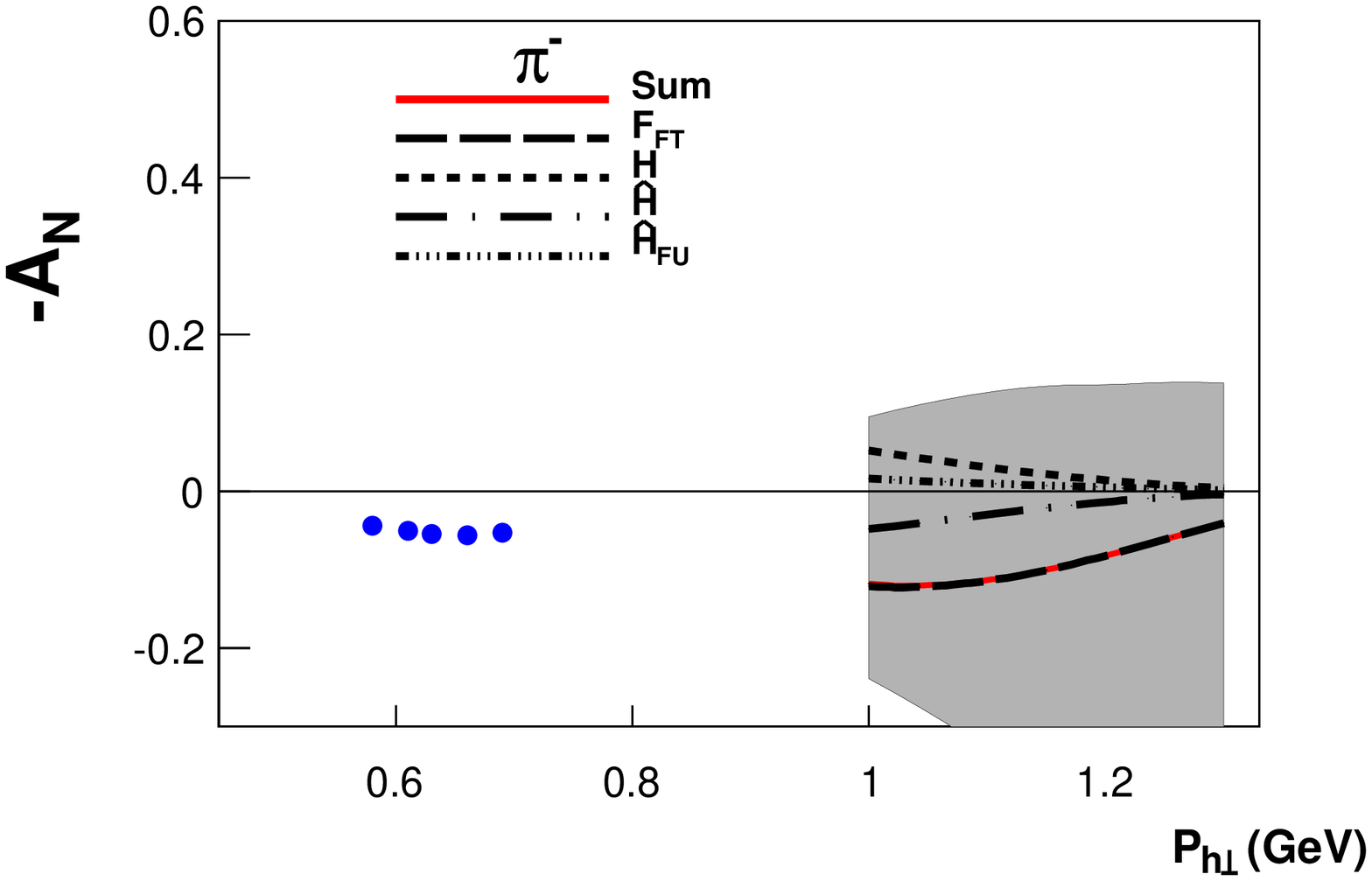}
\caption{$-A_N$ as function of $\Php$ for $\pi^+$ (left panel) and $\pi^-$ (right panel) production off a neutron at $\langle x_F  \rangle \approx -0.26$ and $\sqrt{S} = 3.45$ GeV. 
The data are from  Ref.~\cite{Allada:2013nsw}. 
The description of lines is the same as in Fig~\ref{fig:an_hermes_pipm_xf}.}
\label{fig:an_jlab_pipm_pt}
\end{figure*}
%%%%%%%%%%%%%%%%%%%%%%%%%%%%%%%%%%%

Jefferson Lab published data on $\ell \, N \to h \, X$ collected on a transversely polarized $^3$He target~\cite{Allada:2013nsw}. 
The energy of the experiment is relatively low, such that the largest value of the transverse hadron momentum reached is $\Php = 0.69 \; \rm{GeV}$. 
Therefore, we cannot compare directly to the data. 
However, we can calculate the asymmetry in the region of larger  $\Php$. 
Note that the definition of the reference frame used in Ref.~\cite{Allada:2013nsw} for Jefferson Lab is such that
\bea
A_N (x_F, \Php) = A_{UT}^{\sin{\Psi}}(-x_F^H,\Php) = - A_N^{\rm JLab}(x_F, \Php) \,.
\eea
In Fig.~\ref{fig:an_jlab_pipm_pt} we plot $\pi^\pm$ production on the neutron at JLab 6 for $\Php> 1$ GeV.  
In this case the contribution from the function ${H}^{h/q}$ almost exactly cancels the contributions from $\hat{H}^{h/q}$ and $\hat{H}^{h/q,\Im}_{FU}$, and the asymmetry is close to the result of the contribution $F_{FT}^{q}$. 
 One can see from Fig.~\ref{fig:an_jlab_pipm_pt} that the sign of the asymmetry for both $\pi^+$ and $\pi^-$ is consistent with our calculations, but for $\pi^+$ the trend of the result is much larger than the data.
However, one has to keep in mind that especially for $A_N^{\pi^+}$ the uncertainties of the calculation are quite large where the dominant contribution comes from down quarks, whose Sivers function has rather large errors. 
Future results from JLab 12 \cite{Dudek:2012vr} will allow us to have a better determination of down quark TMDs in the large $x$ region.  

 Extra caution has to be taken when one looks at the comparison of our computations with the experimental data, which seems to show discrepancies. 
Such disagreements can have different sources. 
First, our numerics is based on LO analytical results only.
However, higher order corrections to both the spin-independent and the spin-dependent cross sections can be expected to be very large for the process $\ell \, N \to h\, H$~\cite{Kang:2011jw,Kang:2013lga}, especially in the relatively low $\Php$ region. 
This is actually already confirmed by the HERMES measurement~\cite{Airapetian:2013bim}, where almost all the data correspond to quasi-real photoproduction, and even at the highest $\Php \sim 2$ GeV only a very small fraction of the events satisfies $Q^2 > 1$ GeV$^2$.
Since in collinear factorization quasi-real photoproduction appears for the first time at NLO accuracy, the underlying mechanism of the majority of the data from HERMES (and Jefferson Lab) is not covered by a LO calculation.
In order to obtain a more quantitative understanding of higher order corrections, it would be very useful to have absolute cross section measurements from HERMES and Jefferson Lab. 
At the same time, the NLO calculation has to be carried out in the future.
Along these lines, the positive trend towards larger $\Php$ values in Fig.~\ref{fig:an_hermes_pipm_pt} and, in particular, the relatively good agreement shown in Figs.~\ref{fig:an_hermes_pipm_02z07_xf}, \ref{fig:an_hermes_pipm_z07_xf}, where one has data with $Q^2 > 1$ GeV$^2$, could indeed indicate that issues describing the HERMES data in Figs.~\ref{fig:an_hermes_pipm_xf}, \ref{fig:an_hermes_pipm_pt} may be attributed to possibly large radiative corrections due to those data being at $Q^2 \sim 0$ GeV$^2$.
Second, recall that the error bands are underestimated (see the discussion before Sec.~III B), and may actually overlap the data once fully calculated.
% ===========================
\subsection{Predictions}
\begin{figure*}[hbt]
\centering
      \includegraphics[scale=0.425]{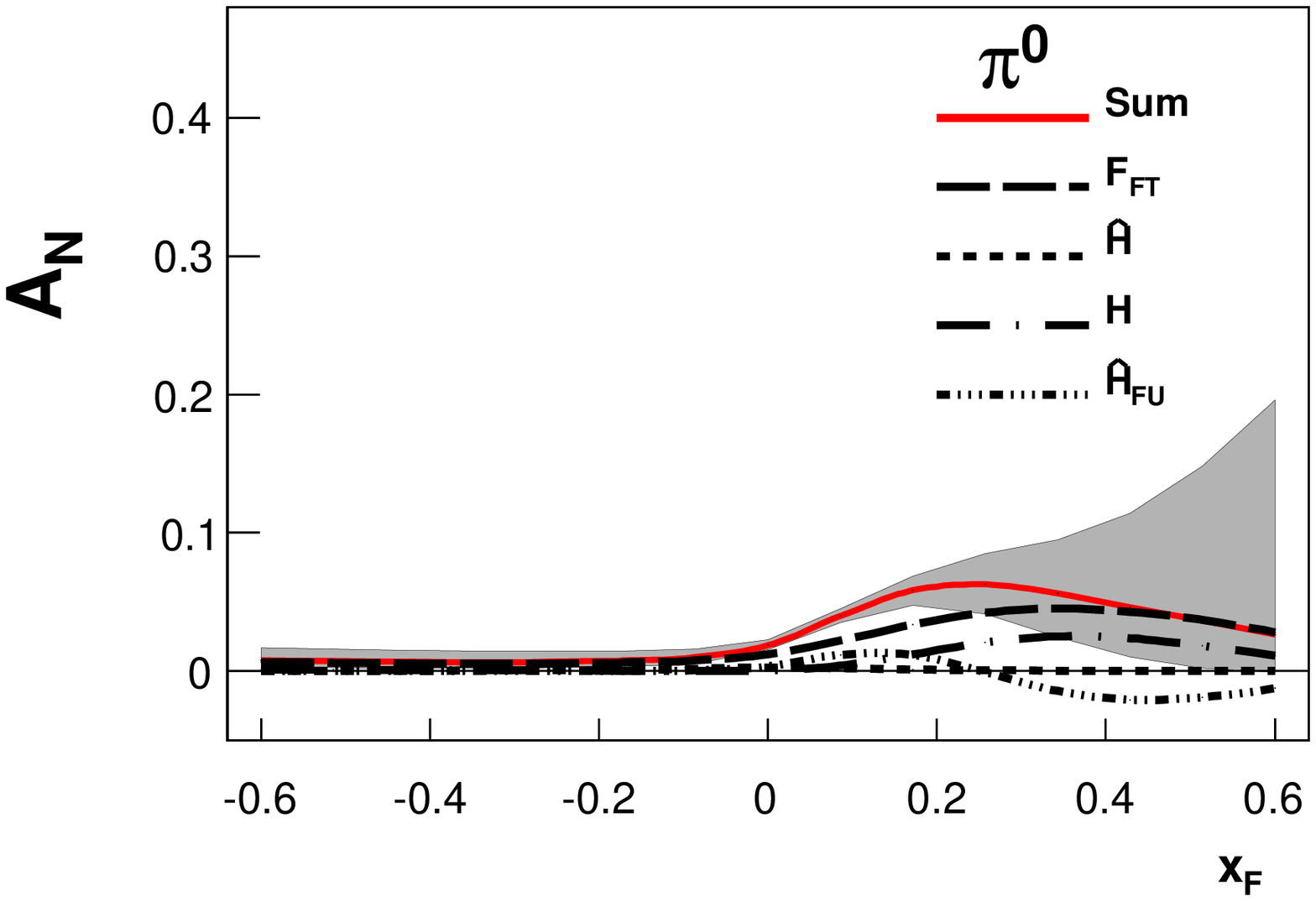} \hspace{-0.7cm}
      \includegraphics[scale=0.425]{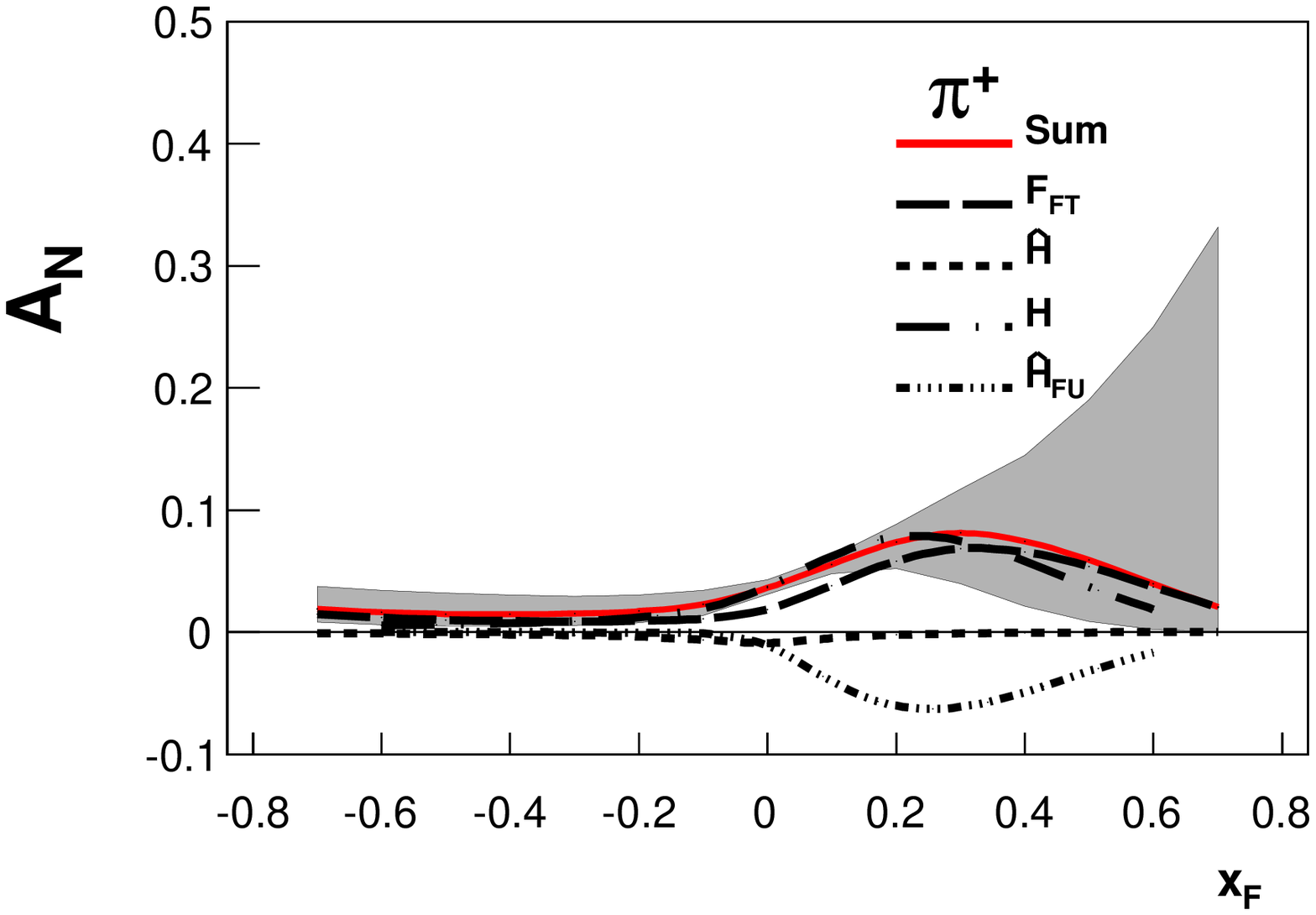}
\caption{Prediction for $A_N$ as function of $x_F$ for $\pi^0$  (left panel) and $\pi^+$ (right panel) production at $\Php = 3 \; \rm{GeV}$ for EIC kinematics ($\sqrt{S} = 63 \; \rm{GeV}$). 
The description of lines is the same as in Fig~\ref{fig:an_hermes_pipm_xf}.}
\label{fig:an_eic_pi0_xf}
\end{figure*}

\begin{figure*}[hbt]
\centering
      \includegraphics[scale=0.425]{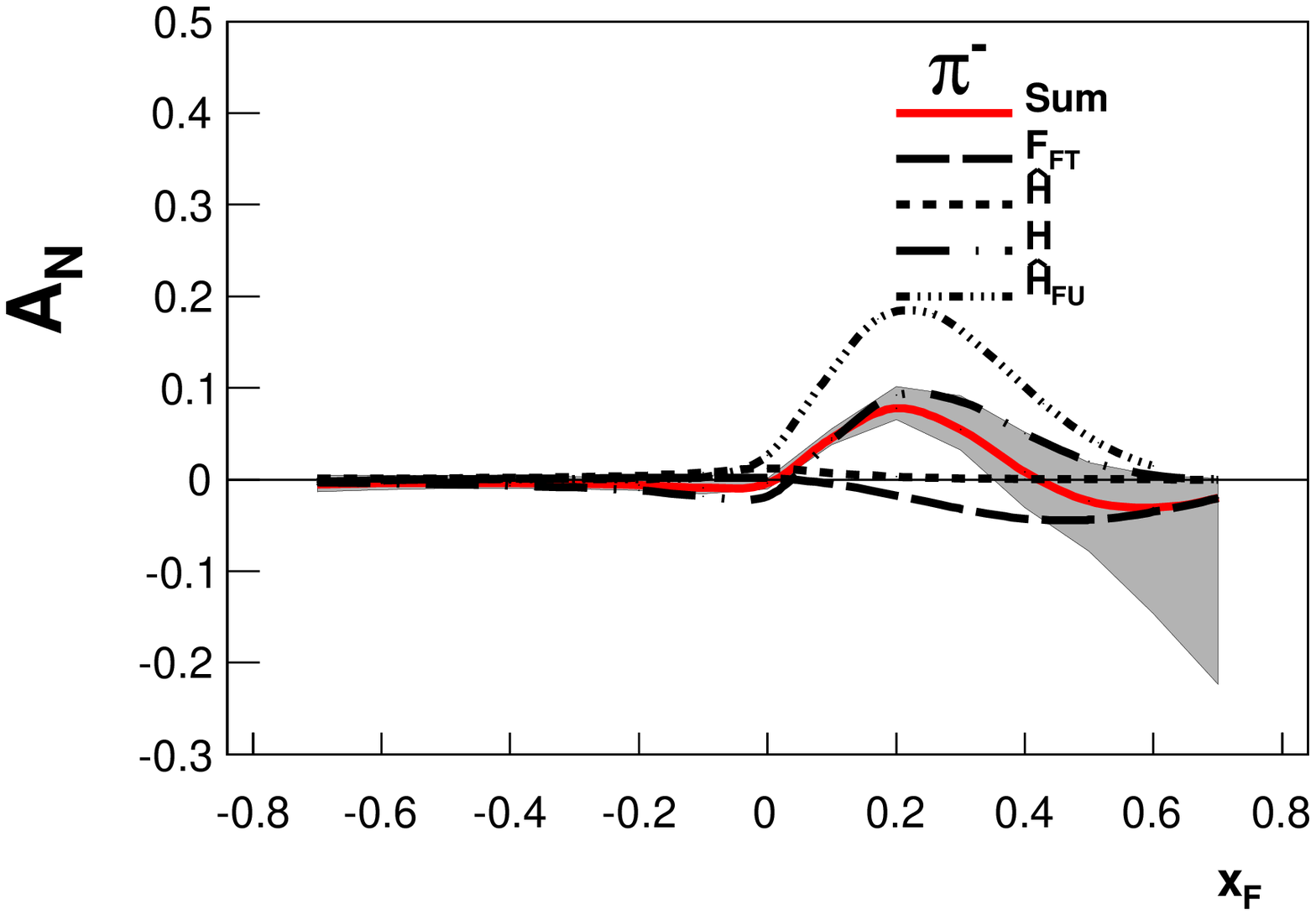}\hspace{-0.7cm}
      \includegraphics[scale=0.425]{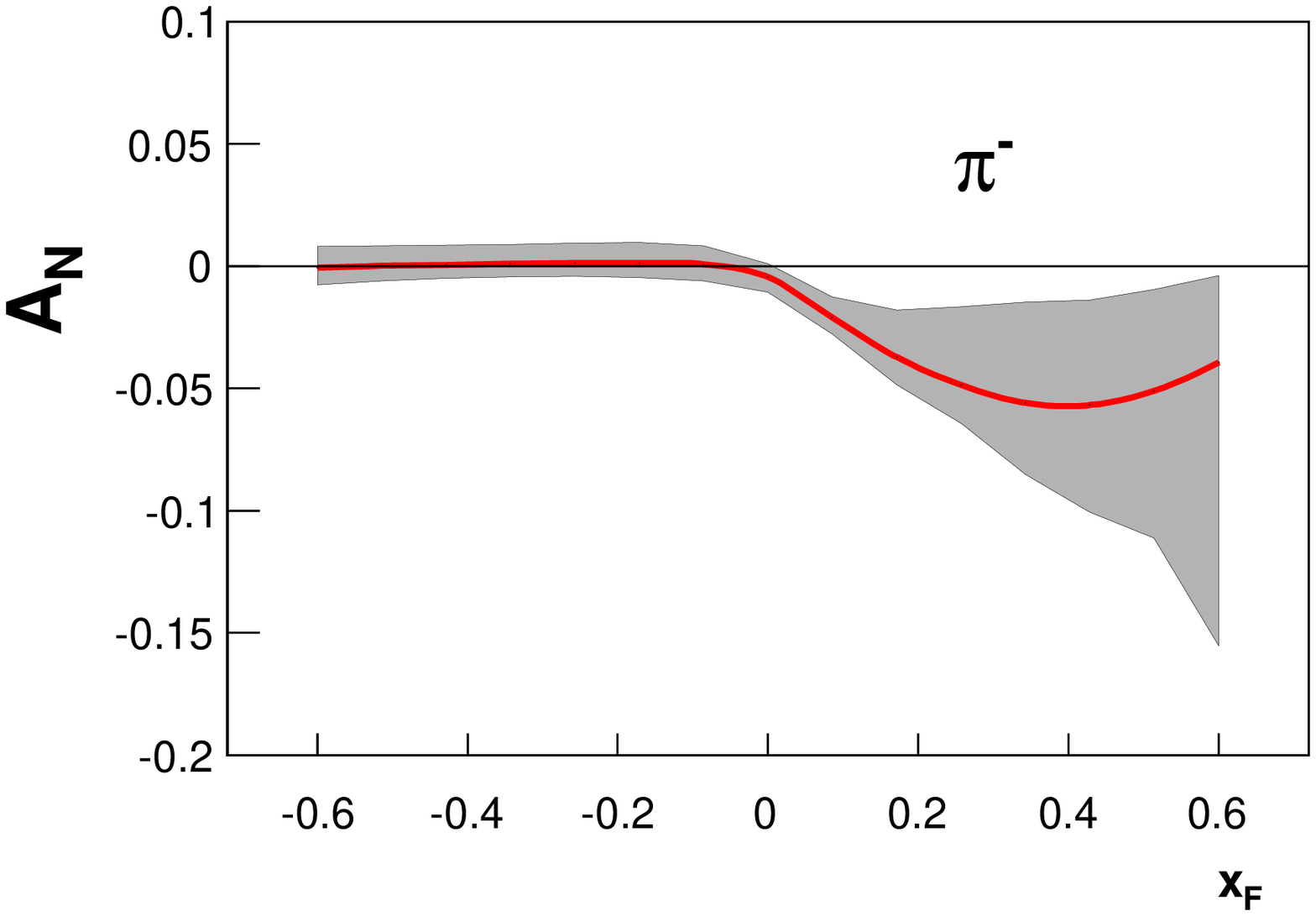}
\caption{Prediction for $A_N$ as function of $x_F$ for $\pi^-$ (left panel) and $\pi^-$ (right panel with $\hat{H}^{h/q,\Im}_{FU} = 0$) production at $\Php = 3 \; \rm{GeV}$ for EIC kinematics ($\sqrt{S} = 63 \; \rm{GeV}$). 
The description of lines is the same as in Fig~\ref{fig:an_hermes_pipm_xf}.}
\label{fig:an_eic_pipm_xf}
\end{figure*}

\begin{figure*}[hbt]
\centering
      \includegraphics[scale=0.425]{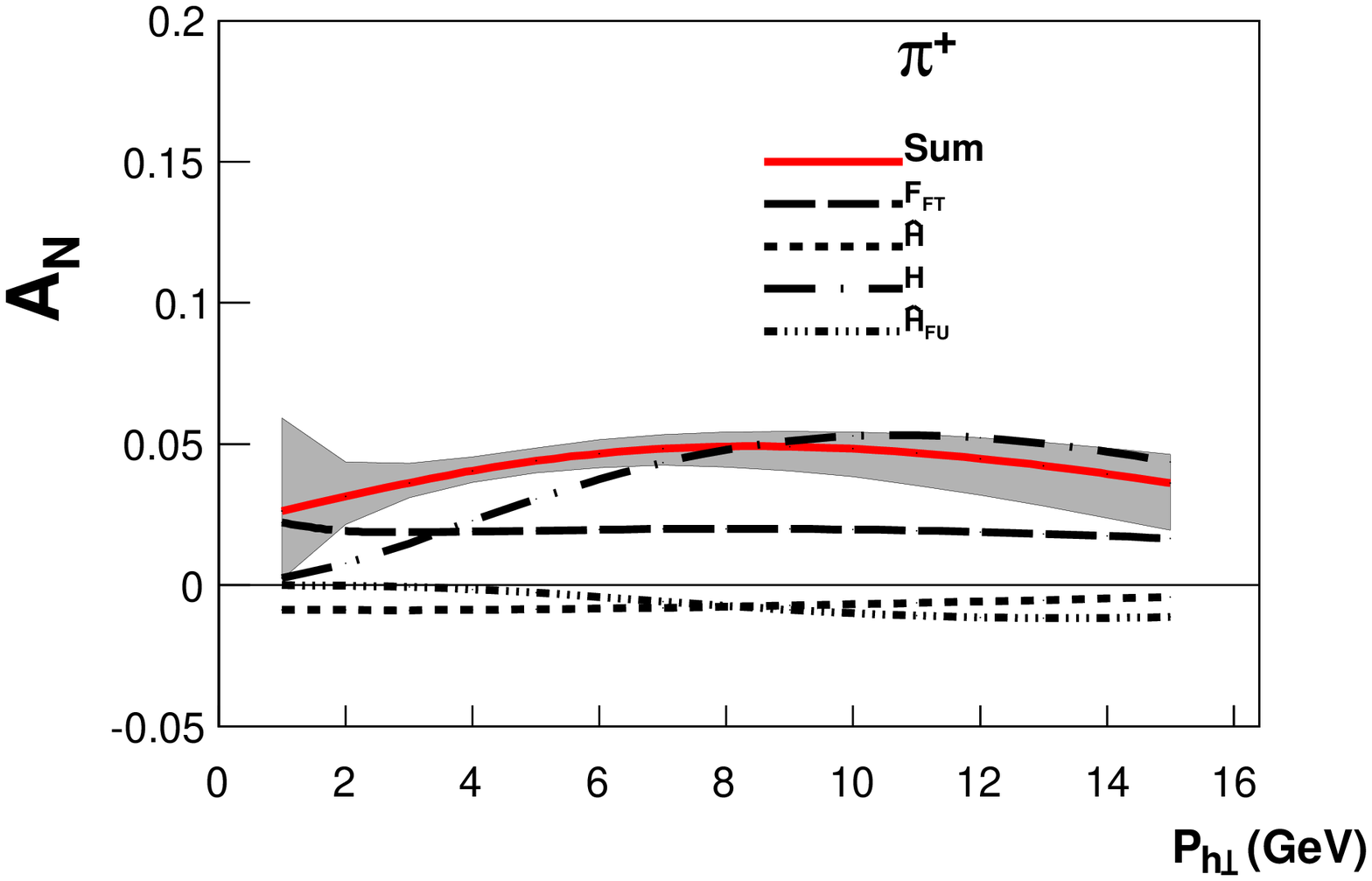}\hspace{-0.7cm}
      \includegraphics[scale=0.425]{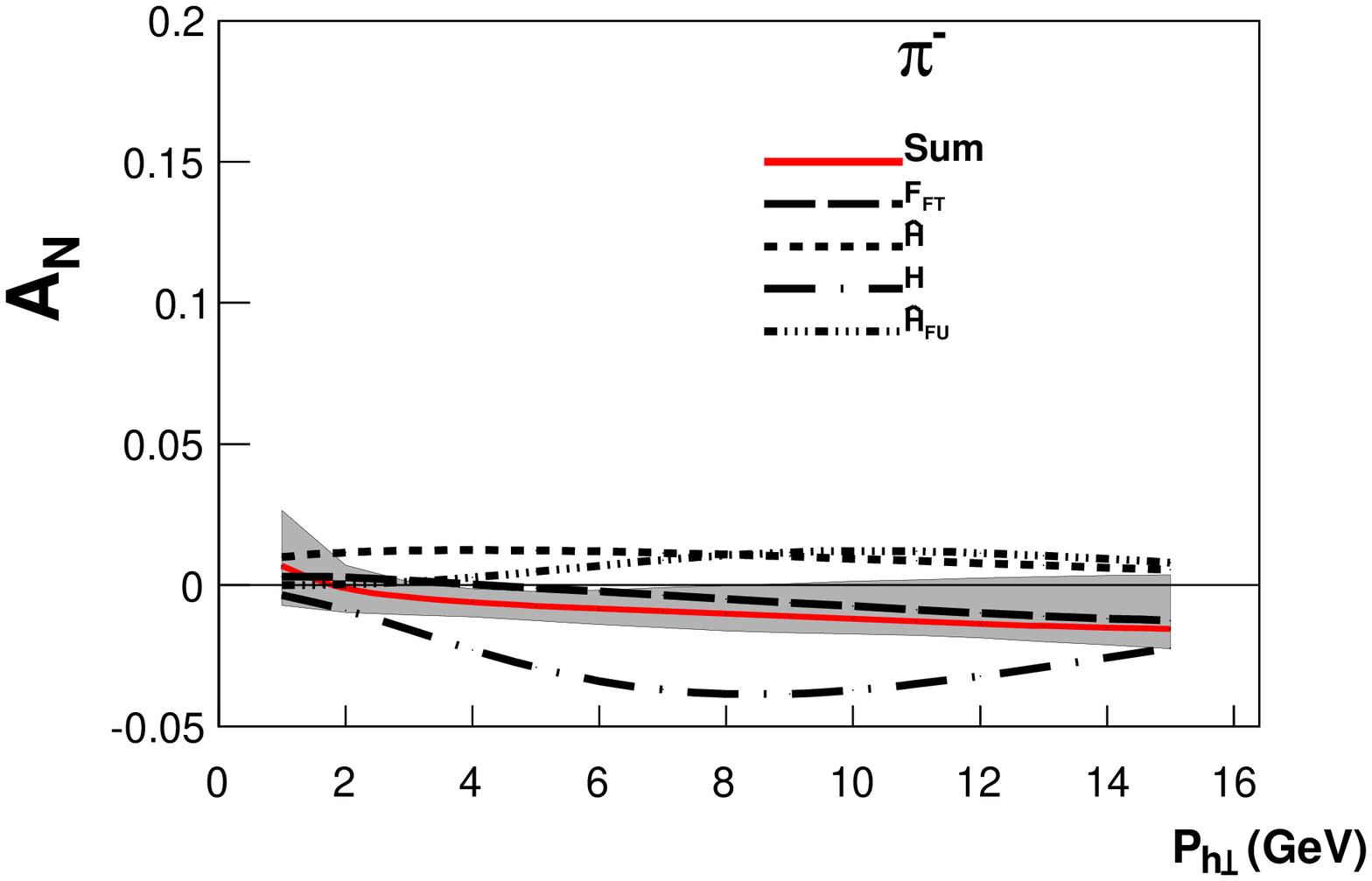}
\caption{Prediction for $A_N$ as function of $\Php$ for $\pi^+$ (left panel) and $\pi^-$ (right panel) production at $x_F = 0$ for EIC kinematics ($\sqrt{S} = 63 \; \rm{GeV}$). 
The description of lines is the same as in Fig~\ref{fig:an_hermes_pipm_xf}.}
\label{fig:an_eic_pipm_pt}
\end{figure*}

\begin{figure*}[hbt]
\centering
      \includegraphics[scale=0.425]{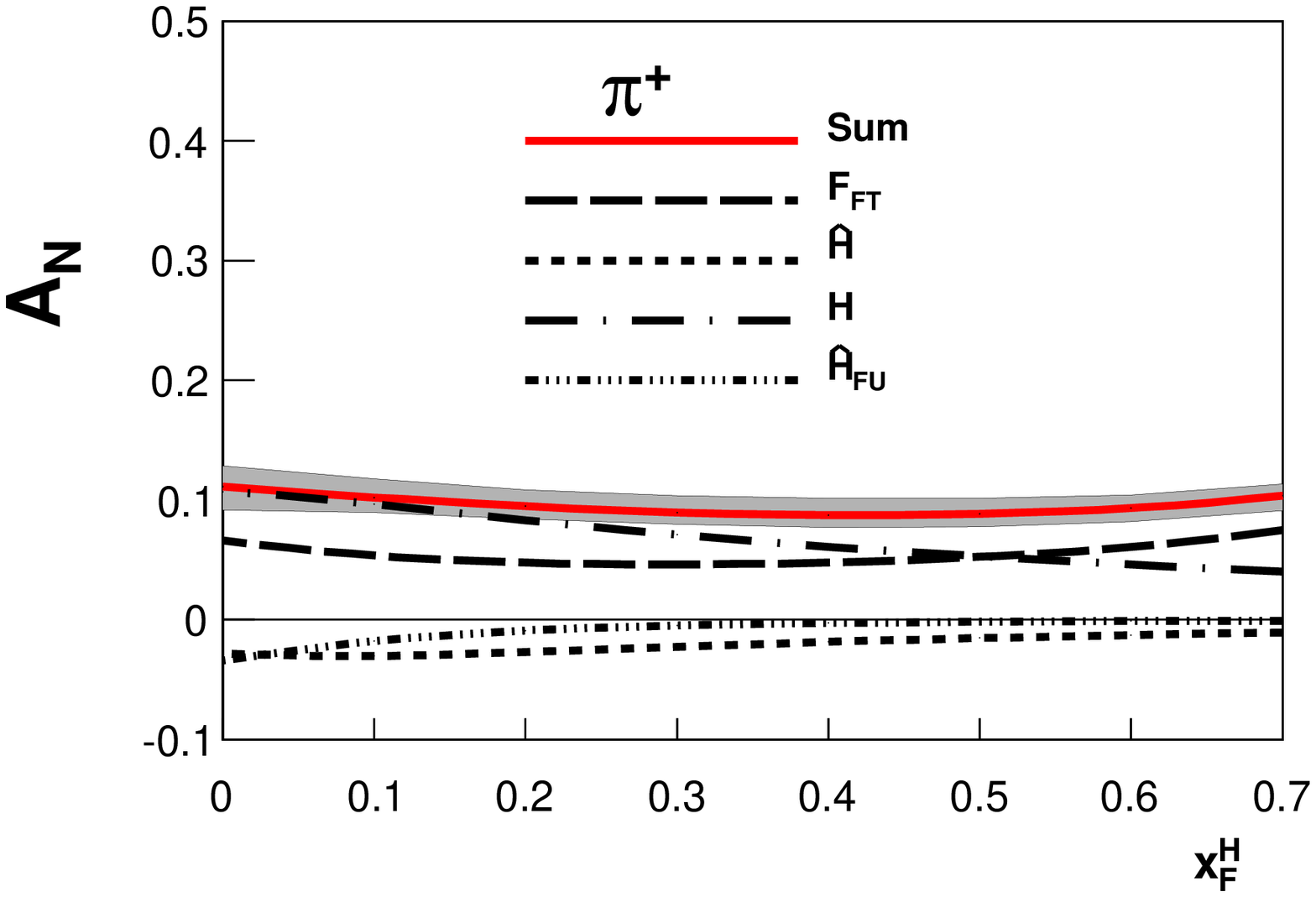}\hspace{-0.7cm}
      \includegraphics[scale=0.425]{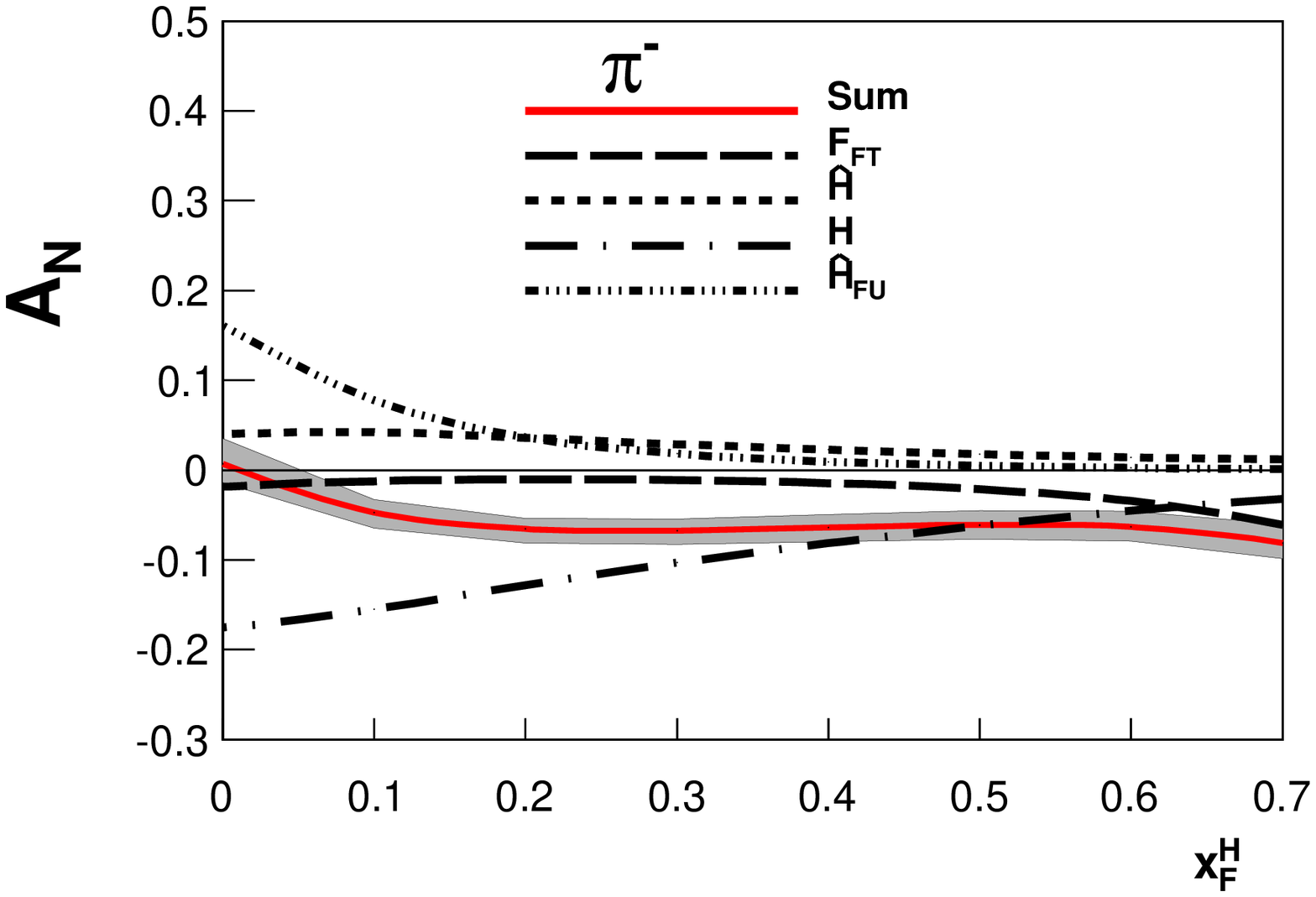}
\caption{Prediction for $A_N$ as function of $x_F^H$ for $\pi^+$ (left panel) and $\pi^-$ (right panel) production at $\Php = 2 \; \rm{GeV}$ for COMPASS kinematics ($\sqrt{S} = 17.3 \; \rm{GeV}$). 
The description of lines is the same as in Fig~\ref{fig:an_hermes_pipm_xf}.}
\label{fig:an_compass_pipm_xf}
\end{figure*}

\begin{figure*}[hbt]
\centering
      \includegraphics[scale=0.425]{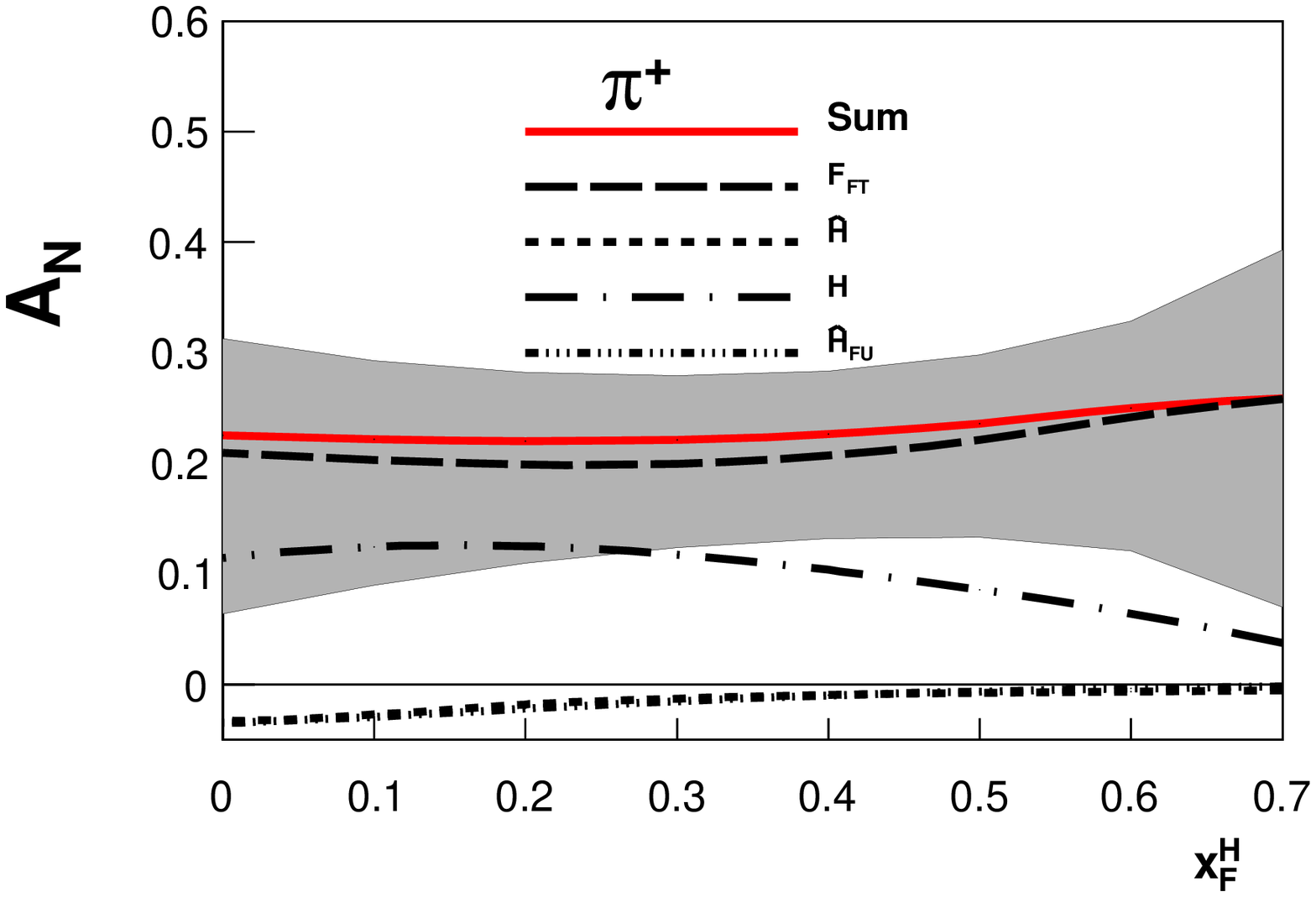}\hspace{-0.7cm}
      \includegraphics[scale=0.425]{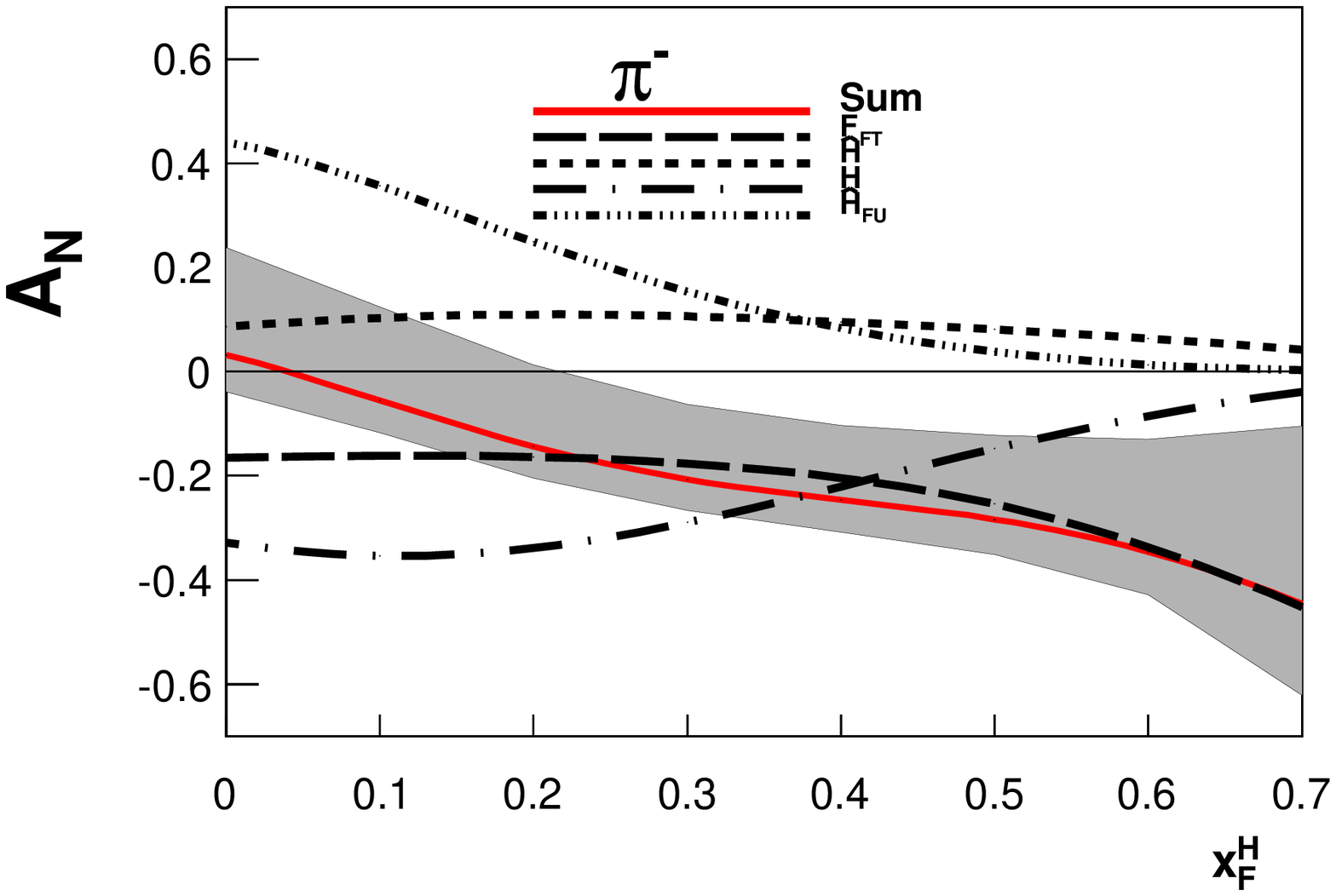}
\caption{Prediction for $A_N$ as function of $x_F^H$ for $\pi^+$ (left panel) and $\pi^-$ (right panel) production at $\Php = 1 \; \rm{GeV}$ for JLab 12 kinematics ($\sqrt{S} = 4.6 \; \rm{GeV}$). 
The description of lines is the same as in Fig~\ref{fig:an_hermes_pipm_xf}.}
\label{fig:an_jlab12_pipm_xf}
\end{figure*}

In this subsection, we show predictions for $A_N$ in the kinematics relevant to several upcoming/planned experiments. 
A future EIC~\cite{Boer:2011fh,Accardi:2011mz,Accardi:2012qut} with variable energy $\sqrt{S} = 50 - 100 \; \rm{GeV}$ will be an ideal facility to study inclusive hadron production in $\ell \, p^\uparrow \to h \, X$.
One reason is the possibility to measure at (much) larger values of $\Php$ where the theory for this process should be under better control.  
We plot in Figs.~\ref{fig:an_eic_pi0_xf},~\ref{fig:an_eic_pipm_xf} our predictions for $\pi^0,\pi^+,\pi^-$  production at $\sqrt{S} = 63 \; \rm{GeV}$ and $\Php = 3 \; \rm{GeV}$. 
Note that for $p^{\uparrow} p \to \pi \, X$ in the forward region ($x_F > 0$) very large values for $A_N$ have been observed.  
We find that a clearly non-zero $A_N$ is predicted in this region. 
An EIC would be in a unique position to make a measurement for $x_F > 0$. 
As already alluded to in the discussion of Fig.~\ref{fig:an_hermes_HF_pipm_xf}, $\pi^-$ production would be particularly interesting in order to study the underlying mechanism of $A_N$. 
One sees that setting $\hat{H}^{h/q,\Im}_{FU} = 0$, as in right panel of Fig.~\ref{fig:an_eic_pipm_xf}, leads to a negative $A_N$, opposite in sign for small to moderate $x_F$ to the case where one keeps $\hat{H}^{h/q,\Im}_{FU}$ nonzero.  
Therefore, a measurement of $A_N$ at $x_F > 0$ at an EIC can help constrain/test the extraction of this 3-parton fragmentation function performed in Ref.~\cite{Kanazawa:2014dca}.
Predictions of $A_N$ as a function of $\Php$ at $x_F = 0$ for $\pi^+,\pi^-$ production are shown in Fig.~\ref{fig:an_eic_pipm_pt}.  
One finds a rather flat $P_{h\perp}$ dependence like in the $pp$ case \cite{Kanazawa:2014dca}. 
In Fig.~\ref{fig:an_compass_pipm_xf} we present our predictions for $A_N$ as a function of $x_F$ at $\Php = 2 \; \rm{GeV}$ for both $\pi^+$ and $\pi^-$ production for the COMPASS experiment at $\sqrt{S} = 17.3 \; \rm{GeV}$. 
Similar predictions at $\Php = 1 \; \rm{GeV}$ are shown in Fig.~\ref{fig:an_jlab12_pipm_xf} for JLab 12 at $\sqrt{S} = 4.6 \; \rm{GeV}$. 
It will be interesting to have experimental data on $A_N$ from all these facilities in the future.

%%%%%%%%%%%%%%%%%%%%%%%%%%%%%%%%%%%
%%%%%%%%%%%%%%%%%%%%%%%%%%%%%%%%%%%

%%%%%%%%%%%%%%%%%%%%%%%%%%%%%%%%%%%
%%%%%%%%%%%%%%%%%%%%%%%%%%%%%%%%%%%

%%%%%%%%%%%%%%%%%%%%%%%%%%%%%%%%%%%
%%%%%%%%%%%%%%%%%%%%%%%%%%%%%%%%%%%

%%%%%%%%%%%%%%%%%%%%%%%%%%%%%%%%%%%
%%%%%%%%%%%%%%%%%%%%%%%%%%%%%%%%%%%

%%%%%%%%%%%%%%%%%%%%%%%%%%%%%%%%%%%
%%%%%%%%%%%%%%%%%%%%%%%%%%%%%%%%%%%

% ===========================
\subsection{Comparison with the Generalized Parton Model}

Here we give a brief  comparison between the collinear twist-3 approach and the GPM from both a conceptual and a phenomenological point of view.
The GPM has been applied to $A_N$ for $\ell \, p^{\uparrow} \to h \, X$~\cite{Anselmino:1999gd,Anselmino:2009pn,Anselmino:2014eza} and for $p^{\uparrow} p \to h \, X$ --- see~\cite{Anselmino:1994tv,Anselmino:1998yz,Anselmino:2005sh,Anselmino:2012rq,Anselmino:2013rya} and references therein.
This model uses 2-parton correlation functions only, but consistently keeps the transverse parton momenta at all stages of the calculation.
(This procedure may lead to a singularity for processes like $p \, p \to h \, X$ upon integrating over transverse momenta, which can however be avoided by introducing an integration cutoff \cite{D'Alesio:2004up,Wang:1998ww,Zhang:2001ce}.) In the case of twist-3 observables like $A_N$ not all leading power terms are covered by the GPM.\footnote{A closely related discussion about the twist-3 so-called Cahn effect in SIDIS can be found in Ref.~\cite{Bacchetta:2008xw}.}
This holds for the twist-3 effect on the distribution side~\cite{Gamberg:2010tj} and, in particular, also for the twist-3 fragmentation contribution~\cite{Metz:2012ct}. 
As mentioned above, for the latter one has two independent fragmentation correlators~\cite{Metz:2012ct}, while in the GPM only the Collins function contributes.
(At present, a detailed analytical comparison of the fragmentation contributions in the two approaches does not exist.)
On the other hand, the GPM contains certain (kinematical) higher twist contributions and may also mimic effects of a collinear higher order calculation at leading twist. 
We note in passing that a recipe for incorporating in the GPM the process dependence of the Sivers effect~\cite{Collins:2002kn} has been discussed in~\cite{Gamberg:2010tj}.

Let us now turn to the phenomenology of $A_N$ for $\ell \, p^{\uparrow} \to h \, X$.
The GPM predictions are closer to the HERMES data than what we found in the collinear twist-3 framework, where the best results in the GPM were obtained by exploiting somewhat older extractions of the Sivers function and the Collins function~\cite{Anselmino:2005ea,Anselmino:2007fs} --- compare Fig.~1 and Fig.~2 in~\cite{Anselmino:2014eza} with our Fig.~\ref{fig:an_hermes_pipm_xf}.
However, one again has to keep in mind the aforementioned underestimated error of the twist-3 calculation and the need for a NLO calculation.
Moreover, due to large error bands, no conclusion could be drawn as to whether the Sivers or Collins effect can describe $A_N$ in $p^\uparrow p \to \pi X$ within the GPM~\cite{Anselmino:2012rq,Anselmino:2013rya}. 
In this regard, a much more definite statement was made with the collinear twist-3 analysis performed in Ref.~\cite{Kanazawa:2014dca}, i.e., that the fragmentation mechanism in that formalism can be the cause of the transverse single-spin asymmetries seen in pion production from proton-proton collisions.  

We find that our results with $\hat{H}^{h/q,\Im}_{FU} = 0$ have the same signs and are close in magnitude to the curves labeled as SIDIS 2 in Figs.~1 and 2 of Ref.~\cite{Anselmino:2014eza} for $\pi^+$ and $\pi^-$ production, respectively. 
One may speculate then that an analytical relation between the GPM and twist-3 approaches (showing where the two formalisms agree and/or differ) is perhaps possible for this observable if one neglects the 3-parton FF.  
However, as already stated, no such rigorous derivation has been performed yet.
Let us also mention that our prediction for $A_N^{\pi^+}$ for the EIC in Figs.~\ref {fig:an_eic_pipm_xf},~\ref{fig:an_eic_pipm_pt} are comparable both in sign and size with those of Refs.~\cite{Anselmino:2008sga,Anselmino:2014eza} using GPM framework.  On the other hand, our result for $A_N^{\pi^-}$ for the EIC is quite different from what one finds in the GPM~\cite{Anselmino:2008sga,Anselmino:2014eza}. 
Such a measurement might therefore allow one to discriminate between the phenomenology of the two approaches. 
 
%
% 4. Section: Summary and discussion
% ===========================
%
\section{Summary \label{sectionIV}}
\noindent
Within the collinear twist-3 factorization formalism, we derived LO results for the transverse single-spin asymmetry $A_N$ for inclusive electroproduction of hadrons in lepton-nucleon collisions, $\ell \, N^\uparrow \to h \, X$. 
In such a process, $A_N$ receives contributions from the QS function $F_{FT}^{q}$ related to the quark Sivers function, from a twist-3 fragmentation function $\hat{H}^{h/q}$ related to the Collins function, and from two other twist-3 fragmentation functions  $H^{h/q}$ and $\hat{H}_{FU}^{h/q,\Im}$.
We provided numerical estimates for typical kinematics for experiments at HERMES~\cite{Airapetian:2013bim} and at Jefferson Lab~\cite{Allada:2013nsw}, and we compared our results with the HERMES data for $\Php \geq 1 \; \rm{GeV}$.
We found that our theoretical estimates for $A_N$ agree with the HERMES results in sign and roughly in shape, but in terms of magnitude they are typically above the data.
We argued that at present such a discrepancy cannot be considered a failure of the collinear twist-3 formalism. 
We emphasized the need for computing the NLO corrections and assess its impact on $A_N$, especially in the region of lower transverse hadron momenta $\Php$.
Moreover, we explained why the error of our numerical calculations is underestimated.
In this regard it will be important to better constrain the 3-parton fragmentation correlator $\hat{H}_{FU}^{h/q,\Im}$.
On the experimental side, it would be very useful to have absolute cross section measurements from both HERMES and Jefferson Lab, which would help one to obtain a quantitative understanding of the role played by higher order corrections.
We also presented predictions for $A_N$ for Jefferson Lab 12, COMPASS, and a potential future Electron Ion Collider. 
In fact, an EIC would be in a unique position to measure $A_N$ in $\ell p^\uparrow \to hX$ at $x_F > 0$.  
In particular $A_N^{\pi^-}$ might allow one to constrain/test the recent extraction of $\hat{H}^{h/q,\Im}_{FU}$ that can play a crucial role in $A_N$ in $pp$ collisions \cite{Kanazawa:2014dca}, and to discriminate between the GPM and the twist-3 frameworks.
In general, further combined theoretical and experimental efforts will help us to deeper understand the underlying QCD mechanism of transverse single-spin asymmetries.

%
% Acknowledgments
% ===========================
%
\section*{Acknowledgments}
\noindent
This work is supported by the U.S. Department of Energy under Contract No.~DE-FG02-07ER41460 (LG), No.~DE-AC52-06NA25396 (ZK), No.~DE-AC05-06OR23177 (AP), No.~DE-AC02-98CH10886 (DP), the National Science Foundation under Grant No.~PHY-1205942 (AM), as well as the LDRD program at LANL (ZK) and RIKEN BNL Research Center (DP).

% References
% ===========================

\bibliography{ref}
 
\end{document}